\documentclass[11pt]{article}
\usepackage[left=3cm,right=3cm,top=3cm,bottom=3cm]{geometry}
\usepackage{amssymb}
\usepackage{amsfonts}
\usepackage{latexsym}
\usepackage{subfigure}
\usepackage{graphicx}
\usepackage{amsmath}
\usepackage{cite}
\usepackage{caption}
\usepackage[usenames, dvipsnames]{color}
\usepackage{hyperref}
\usepackage{float}
\usepackage{setspace}
\usepackage[utf8]{inputenc}
\hypersetup{colorlinks,linkcolor={blue},citecolor={blue},urlcolor={red}}
\usepackage{longtable}
\usepackage{siunitx}
\usepackage[super]{nth}
\usepackage{xurl}
\usepackage{multirow}
\definecolor{darkgreen}{rgb}{0, 0.5, 0.05}

\newcommand{\deleted}[1]{}

\newcommand{\eqn}[1]{\begin{align}#1\end{align}}
\newcommand{\bs}[1]{\boldsymbol{#1}}

\newcommand{\pare}[1]{\left( #1 \right) }

\newcommand{\fr}[2]{\frac{#1}{#2}}

\def\dd{\mathrm{d}}

\def\bna{\bs{\nabla}}

\def\bbf{\bs{f}}
\def\bF{\bs{F}}

\def\bG{\bs{G}}

\def\bI{\bs{I}}

\def\bM{\bs{M}}

\def\bn{\bs{n}}

\def\bq{\bs{q}}
\def\br{\bs{r}}

\def\bu{\bs{u}}

\def\bv{\bs{v}}

\def\bzero{\bs{0}}

\def\btheta{\bs{\theta}}

\def\bsigma{\bs{\sigma}}
\def\btau{\bs{\tau}}
\def\bomega{\bs{\omega}}

\def\blambda{\bs{\lambda}}

\begin{document}

\title{Settling dynamics of non-Brownian suspension of spherical and cubic particles in Stokes flow}

\author{Dipankar Kundu$^{1}$, Florencio Balboa Usabiaga$^{1,}$\footnote{Email address for correspondence: fbalboa@bcamath.org} and Marco Ellero$^{1,2,3}$\\
{\footnotesize $^1$Basque Center for Applied Mathematics, BCAM , Alameda de Mazarredo 14 , Bilbao 48400, Spain} \\
{\footnotesize $^2$ IKERBASQUE, Basque Foundation for Science, Calle de Maria Dias de Haro 3, 48013, Bilbao,Spain}\\
{\footnotesize $^3$ Complex Fluids Research Group, Department of Chemical Engineering, Faculty of Science and Engineering,  } \\
{\footnotesize Swansea University, Swansea SA1 8EN, United Kingdom}}



\maketitle

\begin{abstract}

The present study investigates the gravity-driven settling dynamics of non-Brownian suspensions consisting of spherical and cubic particles within a triply periodic domain. We numerically examine the impact of solid volume fraction on the evolving microstructure of the suspension using the Rigid MultiBlob method under Stokes flow conditions. Our simulations match macroscopic trends observed in experiments and align well with established semi-empirical correlations across a broad range of volume fractions. At low to moderate solid volume fractions, the settling mechanism is primarily governed by hydrodynamic interactions between the particles and the surrounding fluid. However, frequent collisions between particles in a highly packed space tend to suppress velocity fluctuations at denser regimes. For dilute suspensions, transport properties are predominantly shaped by an anisotropic microstructure; though, this anisotropy diminishes as many-body interactions intensify at higher volume fractions. Notably, cubic particles exhibit lower anisotropy in velocity fluctuations compared to spherical particles, owing to more efficient momentum and energy transfer from the gravity-driven direction to transverse directions.
Finally, bidisperse suspensions with mixed particle shapes show enhanced velocity fluctuations, driven by shape-induced variations in drag and increased hydrodynamic disturbances.
These fluctuations in turn affect the local sedimentation velocity field, leading to the segregation of particles in the mixture.
\end{abstract}

\section{Introduction}

Multiphase flows are found in various industrial, environmental, and biological contexts, where rigid particles, bubbles, or droplets of different shapes and densities are suspended in a continuous phase. The presence of solid particles in a fluid significantly influences overall transport, resulting in complex interactions and behaviors. A classic example of this is sedimentation, which refers to the collective settling of particles due to gravity or centrifugal forces. Sedimentation is crucial in numerous applications, such as slurry transport, separation processes in industrial drainage, red blood cell sedimentation in blood vessels, particle settling in fluidized beds and reactors, and the dynamics of wastewater treatment tanks \cite{o1998coagulation,ji2009experimental,zhong2016cfd,van2008numerical}. Consequently, to understand the coupled behavior of fluid-particle systems in these scenarios, it is essential to resolve spatial and temporal scales that capture the intricate physics of particle interactions and sedimentation.

When a single particle settles in a fluid, the balance between drag and gravitational forces allows it to reach a constant terminal settling velocity. However, as the number of particles or the solid volume fraction in the suspension increases, hydrodynamic interactions between particles hinder individual settling, causing particles to settle at velocities lower than the terminal velocity. The topic of calculating this average, or ``hindered'', settling velocity has received considerable attention in the literature. Theoretical predictions by Batchelor \cite{batchelor1972sedimentation,batchelor1982sedimentation} described the average settling velocity of particles in a suspension, though these predictions are limited to very dilute, statistically homogeneous suspensions under Stokes flow conditions. Beyond the dilute limit, particles interact with multiple neighbors through hydrodynamic forces and collisions, forming dynamic microstructures. These microstructures lead to fluctuations in individual particle velocities around the mean sedimentation velocity, affecting the overall settling behavior. In response, empirical correlations for average settling velocity have been proposed by researchers such as Richardson and Zaki \cite{richardson1954sedimentation} and Garside and Al-Dibouni \cite{garside1977velocity}, based on experimental data across a range of volume fractions. Later, Di Felice \cite{di1999sedimentation} introduced modifications for dilute suspensions settling with inertia, applying a factor less than $1$ to the average settling velocity to account for the effects of particle microstructures.

The configurational distribution of particles in suspension evolves based on the Péclet number ($\text{Pe} = U_{\text{sed}} R / D$), where $U_\text{sed}$ is the characteristic sedimentation velocity, $R$ is the particle size, and $D$ is the colloidal diffusion coefficient.
The Péclet number is the ratio between the characteristic particle diffusive and sedimentation times.
Hamid and Yamamoto \cite{hamid2013direct,hamid2013anisotropic,hamid2013sedimentation} studied the combined effects of hydrodynamic velocity fluctuations and self-diffusion on the sedimentation of monodispersed spherical particles at finite Péclet numbers. Their research showed a clear transition from a regime dominated by Brownian motion to one dominated by hydrodynamic fluctuations. They also found that at $\text{Pe} \geq 30$, diffusivities, velocity fluctuations, and relaxation times display non-Brownian characteristics. Here, we focus on the non-Brownian limit, where the Péclet number is high. In this regime, we track the motion of an ensemble of sedimenting spheres until they reach a steady state, observing changes in microstructure and sedimentation velocity.

In addition to the average settling velocity, interparticle hydrodynamic interactions also create particle velocity fluctuations around the mean settling velocity. Caflisch and Luke \cite{caflisch1985variance} were the first to propose that particle velocity fluctuations diverge as the system size increases. Experimental \cite{nicolai1995particle,nicolai1995effect} and numerical \cite{ladd1996hydrodynamic,ladd1997sedimentation} studies also have shown good agreement in the temporal behavior of these fluctuations. However, Nicolai et al.\ \cite{nicolai1995particle,nicolai1995effect} suggested that velocity fluctuations are independent of system size, while Ladd \cite{ladd1996hydrodynamic,ladd1997sedimentation} found a strong dependence on system size. This discrepancy was later addressed by Segre et al.\ \cite{segre1997long} in an experimental study, introducing the concept of a characteristic correlation length, or ``swirl size''. They observed that velocity fluctuations tend to saturate when the system size exceeds this characteristic swirl size, and they scale as $\phi^{1/3}$ at low volume fractions, in contrast to the $\phi^{1/2}$ scaling suggested by Hinch \cite{hinch1988sedimentation}. Subsequent studies by Cunha et al.\ \cite{cunha2002modeling} and Padding and Louis \cite{padding2008interplay} also supported the $\phi^{1/2}$ scaling at low volume fractions, though the dependence of velocity fluctuations on $\phi$ at higher volume fractions remains unclear. Later, Brenner \cite{brenner1999screening} theoretically predicted that the scaling changes from $\phi^{1/2}$ to $\phi^{1/3}$ at high volume fractions.

Although most research on sedimentation focuses on spherical particles due to their well-understood behavior and the availability of models describing their interaction with fluid flow, many natural and industrial applications, such as environmental pollution control, chemical engineering, and the production of short-fiber composites \cite{li2006effect,tinland2000simultaneous,van2000modelling} involve non-spherical particles. This makes the analysis of such flows more complex. Non-spherical particles can vary in orientation relative to the flow, which affects the forces and torques they experience \cite{holzer2009lattice,zastawny2012derivation} and leads to hydrodynamic behaviors different from those of spherical particles. The sharp edges of anisotropic particles enhance flow separation, increasing the drag coefficient. Furthermore, while the rotation of spherical particles generates minimal vorticity, the rotation of non-spherical particles displaces more fluid and generates significant vorticity. Studies on the free motion of anisotropic particles, such as oblate and prolate spheroids \cite{ardekani2016numerical}, disks \cite{zhong2011experimental,chrust2013numerical}, and cylinder-like objects \cite{mathai2017mass,fernandes2005zigzag} are available in the literature. Given our interest in understanding the effects of particle shape on suspension behavior, we chose cubic particles for this study. Cubes can be defined with a single parameter, similar to spheres, which simplifies the exploration of key governing parameters. Additionally, three orthogonal planes of symmetry of a cube, results in added-mass effects that act solely as resistance to acceleration, without creating cross-coupling between force and torque.

As mentioned earlier, while numerous theoretical, numerical, and experimental studies focus on the sedimentation of spheres, there is a notable lack of similar research on anisotropic particles such as cubes. Mallavajula et al.\ \cite{mallavajula2013intrinsic} conducted simulations and experiments on fluid flow around cubic particles, examining their rheological properties in the dilute regime. Later, Cwalina et al.\ \cite{cwalina2016rheology} investigated the rheology of concentrated cubic colloidal suspensions under steady and dynamic shear flow at low Reynolds numbers ($\text{Re} = \rho U_\text{sed} R / \eta$, where $U_\text{sed}$ is the characteristic sedimentation velocity, $R$ is the particle size, and $\rho$ and $\eta$ are the density and viscosity of the fluid, respectively). They analyzed the steady and dynamic shear viscosities, as well as the difference between the first and second normal stresses, across a wide range of particle concentrations and applied stresses. Rahmani and Wachs \cite{rahmani2014free} conducted a numerical study on the settling behavior of cubic and tetrahedral particles, observing that these particles follow vertical paths, then transition to helical or spiral paths, and eventually exhibit chaotic motion at varying Reynolds numbers. They also demonstrated that the rotation rates and resulting forces on anisotropic particles, such as cubes and tetrahedrons, are much more pronounced compared to those on spheres. Later, Seyed-Ahmadi and Wachs \cite{seyed2019dynamics} attempted to explain the significant variations in the drag coefficient of moving cubes by linking them to the particle's path and vortex structures at different Galileo numbers. Unlike spheres, cubes exhibited helical motion across all density ratios in their study, making it a characteristic type of motion for cubic particles. In a more recent study by the same authors \cite{seyed2021sedimentation}, it was observed that the tendency for columnar cluster formation, measured using particle-pair distribution functions, is somewhat lower for cubes than for spheres at moderate volume fractions with inertia. They attributed this effect to the higher rotation rates of cubes, which lead to an enhanced lift force, increasing the likelihood that a cube will escape from an existing cluster.

Despite extensive research on sedimentation, several open questions remain, such as the nature of velocity fluctuations, the movement of individual particles in suspension, and the formation of microstructures, especially in the context of spherical and cubic particle suspensions. Motivated by these open questions and the intriguing differences between spherical and cubic particles, we are particularly interested in studying how spheres and cubes behave while settling under gravity in the Stokes regime. This study aims to investigate the evolution of particle microstructure and its impact on the transport properties of non-Brownian suspensions of spherical and cubic particles over a wide range of solid volume fractions. We organize our paper as follows: Section \ref{sec:eqs} provides the motivation for the study, describes the simulation setup, and outlines the governing equations and numerical method. Section \ref{section 4} presents the numerical results, including hindered settling velocity, velocity fluctuations, and particle microstructure. Finally, Section \ref{section 5} summarizes the key findings.

\section{Governing equations and numerical method}
\label{sec:eqs}

\begin{figure}[h]
  \centering
  \includegraphics[width=0.9\textwidth]{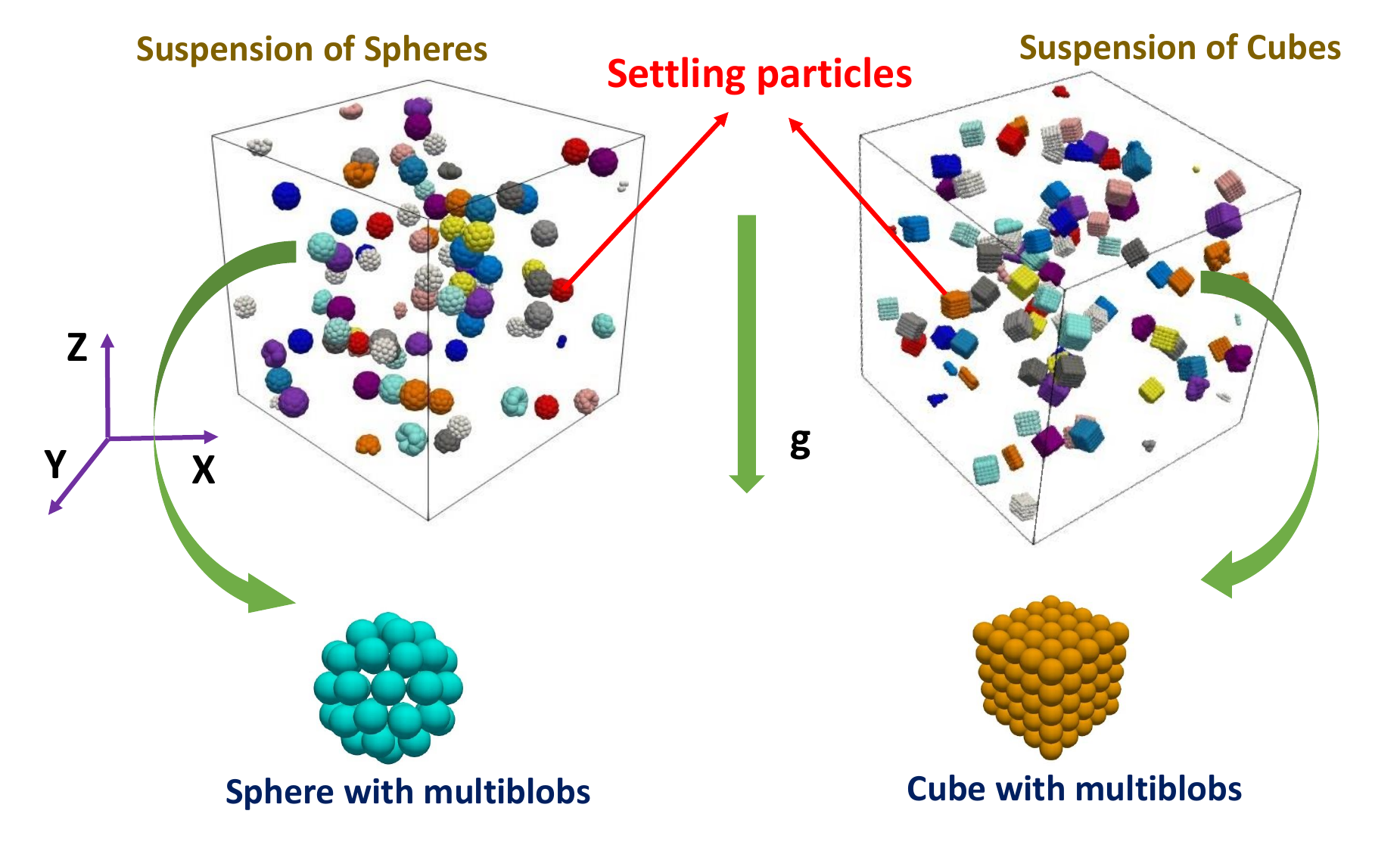}
  \caption{Schematic representation of the fully periodic cubic computational box consisting of a randomly placed spherical/cubic particles.
    The spherical body is discretized with a resolution of $42$ small blobs, whereas a resolution of $125$ blobs is considered for cubic particle.  }
  \label{domain}
\end{figure}
The applied numerical set up is illustrated in Fig.\ \ref{domain}. Two particles of different types are investigated namely: spherical (left) and cubic (right) particles. In simulations, monodisperse spherical/cubic particles are allowed to settle under the action of gravity in a quiescent viscous fluid under creeping flow conditions, i.e.\ $\text{Re}=0$.
The suspension is composed with $M$ rigid particles denoted by $\{\Im_p\}_{p=1}^M$, where each particle is tracked by the position of its center $\bq_p$ and its orientation $\btheta_p$.
We use unit quaternions to represent orientations but other formulations are formally equivalent \cite{Delong2015b}.
The linear and angular velocities of particle $p$ are denoted by $\boldsymbol{u}_p$ and $\boldsymbol{\omega}_p$, respectively.
The total external force applied to particle $p$ is $\bbf_p$, and the total torque around the center of the particle $\btau_p$.
The velocity and pressure, $\boldsymbol{v}$ and $p$, of the fluid with viscosity $\eta$ are governed by the Stokes equations
\begin{align}
-\boldsymbol{\nabla} p + \eta \boldsymbol{\nabla}^2 \boldsymbol{v} = \bzero,  \\
\boldsymbol{\nabla} \cdot \boldsymbol{v} = 0,
\end{align}
while on the particles surfaces the flow obeys the no-slip condition
\begin{equation}
\boldsymbol{v}(r)=\boldsymbol{u}_p+ \boldsymbol{\omega}_p \times (\boldsymbol{r} - \boldsymbol{q}_p) ~~~~~~ \text{for} ~ \boldsymbol{r} \in \partial \Im_p.
\end{equation}
The equations are closed by the balance of force and torque.
The flow exerts a traction on the particles surface $-\blambda = \bsigma \cdot \bn$ where $\bn$ is the surface normal
and $\bsigma = -p\bI + \eta \pare{\bna \bv + (\bna \bv)^T}$ the fluid stress.
This hydrodynamic traction, integrated on the particles surfaces, balance the external forces and torques applied to the particles
\begin{equation}
\begin{aligned}
\int_{\partial \Im_p} \boldsymbol{\lambda}(\boldsymbol{r}) dS_{\boldsymbol{r}}= \boldsymbol{f}_p,  \\
\int_{\partial \Im_p} (\boldsymbol{r} - \boldsymbol{q}_p) \times \boldsymbol{\lambda}(\boldsymbol{r}) dS_{\boldsymbol{r}}= \boldsymbol{\tau}_p .
\label{A_Label}
\end{aligned}
\end{equation}
We construct rigid bodies of essentially arbitrary shape as a collection of rigidly connected "blobs", each having a finite radius $a$ and a position $\boldsymbol{r}_i$, (see Fig.\ \ref{domain}), forming a composite object that we will refer to as a Rigid MultiBlob (RMB).
The hydrodynamic problem is solved by RMB method, which is a regularized boundary integral method, formulated as a symmetric saddle system involving the mobility matrix between markers.
A detailed description of this method can be found in Refs. \cite{kallemov2016immersed, balboa2017hydrodynamics}.
After discretizing the particles, the integrals in the balance of force and torque (eq.~\eqref{A_Label}) become sums over the blobs, while the no-slip condition is applied on each blob
\begin{equation}
\begin{split}
\sum_{\boldsymbol{r}_i \in \Im_p} \boldsymbol{\lambda}_i &= \boldsymbol{f}_p,   \\
\sum_{\boldsymbol{r}_i \in \Im_p} (\boldsymbol{r}_i - \boldsymbol{q}_p) \times \boldsymbol{\lambda}_i &= \boldsymbol{\tau}_p,  \\
\bv\pare{\br_i} = \boldsymbol{u}_p+ \boldsymbol{\omega}_p \times (\boldsymbol{r}_i - \boldsymbol{q}_p) &= \sum_{j} \boldsymbol{M}_{ij} \boldsymbol{\lambda}_j ~~~~~ \text{for} ~ \boldsymbol{r}_i \in \Im_p.
\end{split}
\label{B_Label}
\end{equation}
In the no-slip equation $\blambda_j$ represents a finite force acting on the blob $j$ while
the mobility matrix $\boldsymbol{M}_{ij}$ couples the force acting on the blob $j$ to the velocity of blob $i$.
The mobility $\boldsymbol{M}_{ij}$ is constructed by a double convolution of the Greens' function of the Stokes equation, $\bG_{\text{St}}(\br, \br')$,
with two Dirac delta functions defined on spheres of radius $a$ centered in the blobs $i$ and $j$ \cite{Rotne1969, Wajnryb2013}
\eqn{
  \label{eq:RPY}
  \bM_{ij} = \bM(\br_i, \br_j)  &=  \fr{1}{(4\pi a^2)^2}  \int \delta(|\br' - \br_i|-a) \bG_{\text{St}}(\br', \br'') \delta(|\br'' - \br_j|-a) \dd^3 r'' \dd^3 r'.
}
This regularization of the Green's function makes the mobility $\bM$ positive definite and the numerical method quite robust and easy to implement.
To include the effect of the periodic boundary condition (PBC), we compute the action of the mobility matrix with a Fast Multipole Method that accounts for the periodic images
with a linear computational cost \cite{Yan2020}.

The mobility problem \eqref{B_Label} is a linear system for the unknown particles velocities, $\bu_p$ and $\bomega_p$, and the force acting on the blobs, $\blambda_i$.
The structure of the equations is more evident if we rewrite them in a compact matrix form.
To do so, first, we define composite variables with the velocities $\boldsymbol{U}_p = \{\boldsymbol{u}_p, \boldsymbol{\omega}_p \}$
and the external forces-torques $\boldsymbol{F}_p = \{\boldsymbol{f}_p, \boldsymbol{\tau}_p \}$ of particle $p$,
and the whole system of particles $\boldsymbol{U}=\{\boldsymbol{U}_p\}_{p=1}^M$ and $\bF=\{\bF_p\}_{p=1}^M$.
With this notation we can write the mobility problem \eqref{B_Label} as the saddle-point linear system of equations
\begin{equation}
  \label{eq:linear_system}
\begin{bmatrix}
\boldsymbol{M} & -\boldsymbol{K} \\
-\boldsymbol{K^T} & \boldsymbol{0}
\end{bmatrix}
\begin{bmatrix}
\boldsymbol{\lambda}  \\
\boldsymbol{U} 
\end{bmatrix}
=
\begin{bmatrix}
\boldsymbol{0}  \\
\boldsymbol{-F} 
\end{bmatrix},
\end{equation}
where $\boldsymbol{K}$ is a block matrix that converts the particles velocities into blobs velocities and its transpose sums over the force on the blobs to give the total force-torque acting on the particles. For a many particle suspension \eqref{eq:linear_system} is a large linear system than can be solved efficiently by a preconditioned GMRES iterative solver \cite{balboa2017hydrodynamics}.
Finally, the spheres and cubes are pulled by an external force along the $z$-axis to model the pull of gravity.
With the finite resolution used to discretize particles, see Fig.\ \ref{domain}, lubrication forces are not recovered exactly,
therefore, an additional repulsive force is incorporated to account for the steric interactions between solid particles,
effectively preventing artificial overlaps.
The force, acting between blobs forming the colloids, is derived from the potential
\[
    \psi(\br)= 
\begin{cases}
    \psi_0 \exp \left(-\frac{r-2a}{b} \right) & \text{if } r > 2a, \\
    \psi_0 + \psi_0 \fr{2a - r}{b}              & \text{if } r \leq 2a,
\end{cases}
\]
where $\psi_0$ denotes the strength of the repulsion, while $b$ characterizes the decay length of the interaction.


\section{Results and Discussion}
\label{section 4}


\begin{longtable}{llcc}
  \caption{Physical parameters used in the present study:
  } \\  
  \hline
  Symbol & Description & Value & Comments \\ 
  \hline
  \endhead
  $M$ & No.\ of particles & $125$ & unless otherwise state \\
  $R$ & Particle radius &  $1$ & for cubes the volume-equivalent sphere \\
  & & & radius is used, the cube side is \\
  & & & $h=(4\pi/3)^{1/3} R$ \\
  $r_s$ & Blob radius for sphere &  $0.2435$  & spheres are discretized with 42 blobs \\
  $r_c$ & Blob radius for cube &   $0.1746$  & cubes are discretized with 125 blobs \\
  $\phi$ & Solid volume fraction & $0.001$ \text{to} $0.5$   \\
  $\eta$ & Fluid viscosity & $1$  \\
  $f$ & Gravity force per particle & $1$  \\
  $\psi_0$ & Repulsion strength & $0.011$ & for spheres and cubes \\
  $b$ & Decay length spheres & $r_s / 10$  \\
  $b$ & Decay length cubes & $r_c / 10$  \\  
  \hline
  \label{table:Table 1}
\end{longtable}

Three-dimensional simulations are performed within a cubic, triply periodic domain, illustrated in Fig.\ \ref{domain}.
The computational domain has dimensions of $L \times L \times L$, where $L$ is the length of each side.
To achieve the desired solid volume fraction, $\phi$, we fix the total number of particles, $M$, and adjust the size of the computational box.
The solid volume fraction is then calculated as $\phi=MV/L^3$, where $V$ represents the volume of each particle, either being spherical or cubic in shape.
The characteristic length scale, denoted by $R$, is defined as the radius for spherical particles and the volume-equivalent sphere radius for cubes,
i.e.\ the cubes have side length $h = \pare{4\pi/3}^{1/3} R$.
We use dynamic simulations to obtain the trajectory and velocities of sedimenting particles.
Particles are initially seeded at random positions ensuring that particles do not overlap.
Then, the trajectories are computed for a finite time and ensemble averages are formed over all the particles as time goes on.
In our coordinate system, gravity is aligned in the negative $z$-direction, also referred to as vertical direction in the following.
The horizontal direction corresponds to the $x$ and $y$-coordinates.
The physical parameters and computational settings used in the present study is listed in Table \ref{table:Table 1}.

\subsection{Sedimentation of two spheres}

\begin{figure}[h]
  \centering
  \subfigure[]{\includegraphics[width=0.5\textwidth]{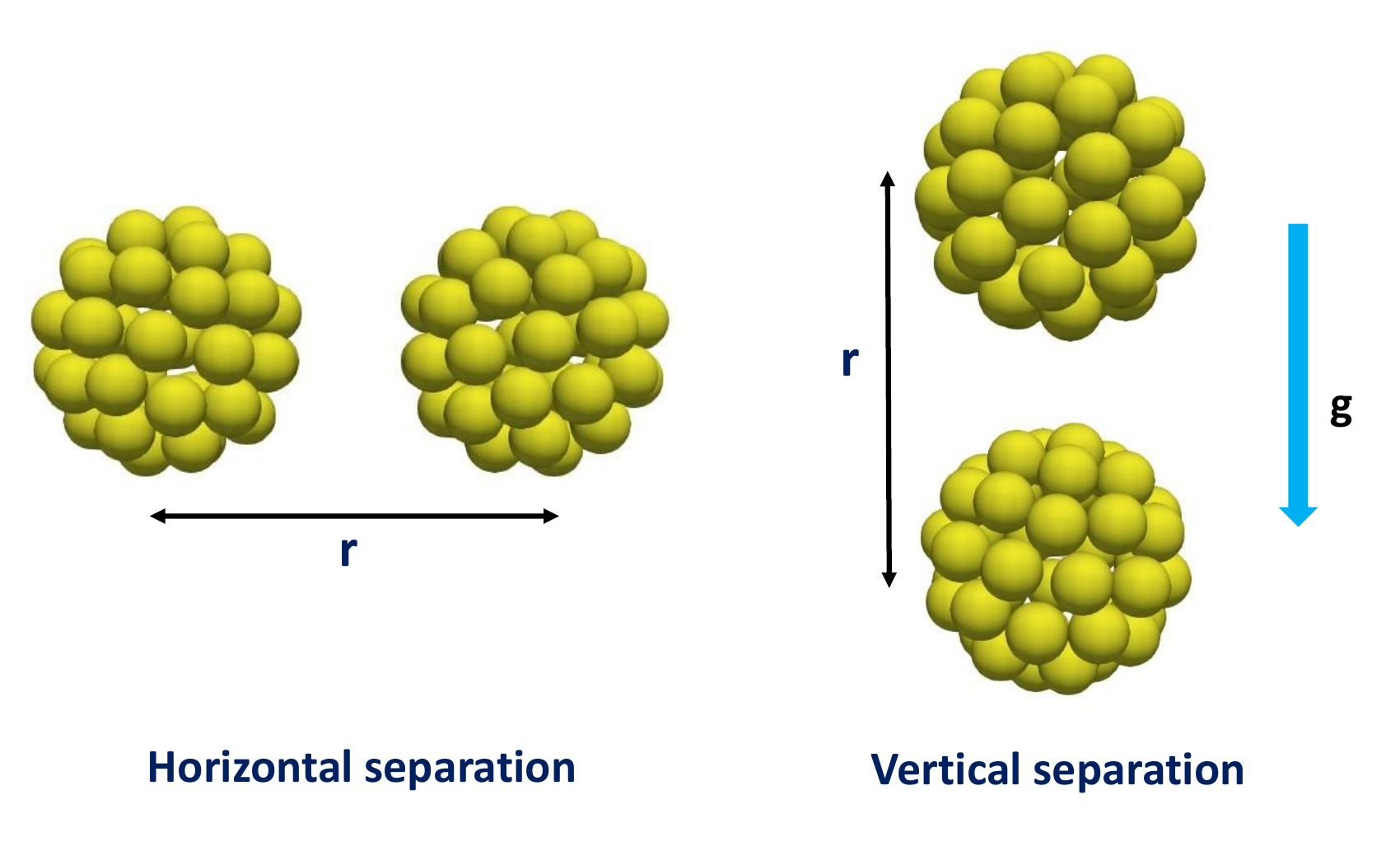}}
  \subfigure[]{\includegraphics[width=0.45\textwidth]{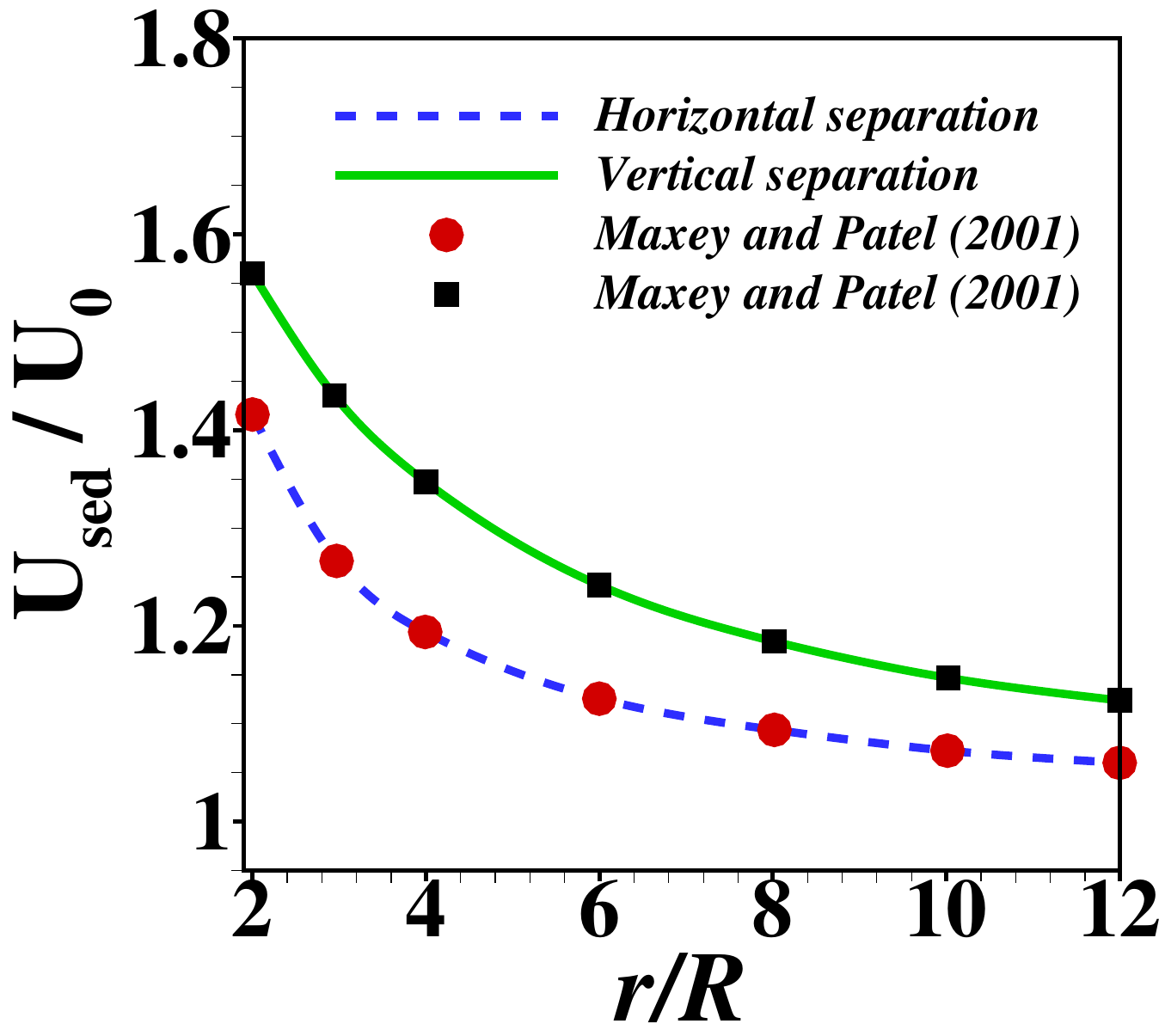}}
  \caption{(a) Schematic representation of two spheres falling freely in an infinite domain, with the spheres initially separated either horizontally or vertically. The separation distance, $r$, refers to the distance between their centers. (b) Comparison of the settling velocities for a pair of equal spheres separated either horizontally or vertically with the result of Maxey and Patel \cite{maxey2001localized}. Here $U_\text{sed}$ and $U_0$ are the settling velocities for a pair of identical spheres and a isolated sphere, respectively. }
\label{compare}
\end{figure}
We begin our investigation by validating the steady motion of a pair of identical and equal sized spheres settling under gravity in unbounded domain (i.e.\ $L \rightarrow \infty $). The initial configuration of two spheres is shown in Fig.~\ref{compare}a. The spheres are either separated horizontally or vertically to the line of center. As pointed out by Batchelor \cite{batchelor1972sedimentation}, the settling velocity depends on the separation between the particles and the orientation of the line of centers to the vertical. Later, Maxey and Patel \cite{maxey2001localized} introduced a finite-valued force multipole expansion to describe the dynamics of spherical particles sedimenting in Stokes flow. They used the Force Coupling Model (FCM) to evaluate the settling velocity of two equal spheres in the above mentioned configurations. If the particles are otherwise isolated and not influenced by other particles or boundaries, they will have equal settling velocities due to the symmetry of the induced Stokes flow. The mutual influence of the disturbance flow created by each particle leads to a settling velocity larger than that for a single isolated particle.
Our numerical result in Fig.\ \ref{compare}b shows a good agreement with the result of Maxey and Patel \cite{maxey2001localized}.
The settling velocity $U_\text{sed}$ is scaled by the Stokes velocity $U_0$ of a single sphere and given in terms of the distance between the centers, so $r/R=2$ corresponds to touching spheres. There is no external torque on the particles and they are free to rotate in response to the local fluid velocity.
Fig.\ \ref{compare}b suggests that the particles settle with the same velocity as an isolated sphere in an infinite medium, when they are very far apart from each other (i.e., $r/R \to \infty$).
However, the mutual influence of the flow disturbance created by each particle results in a settling velocity that is larger than that of a single isolated particle.
This effect leads to a velocity ratio of $U_\text{sed}/U_0 \approx 1.41$ at $r/R = 2.01$ for horizontally separated spheres.

\subsection{Effect of domain size}


\begin{figure}[h]
  \centering
  \subfigure[]{\includegraphics[width=0.45\textwidth]{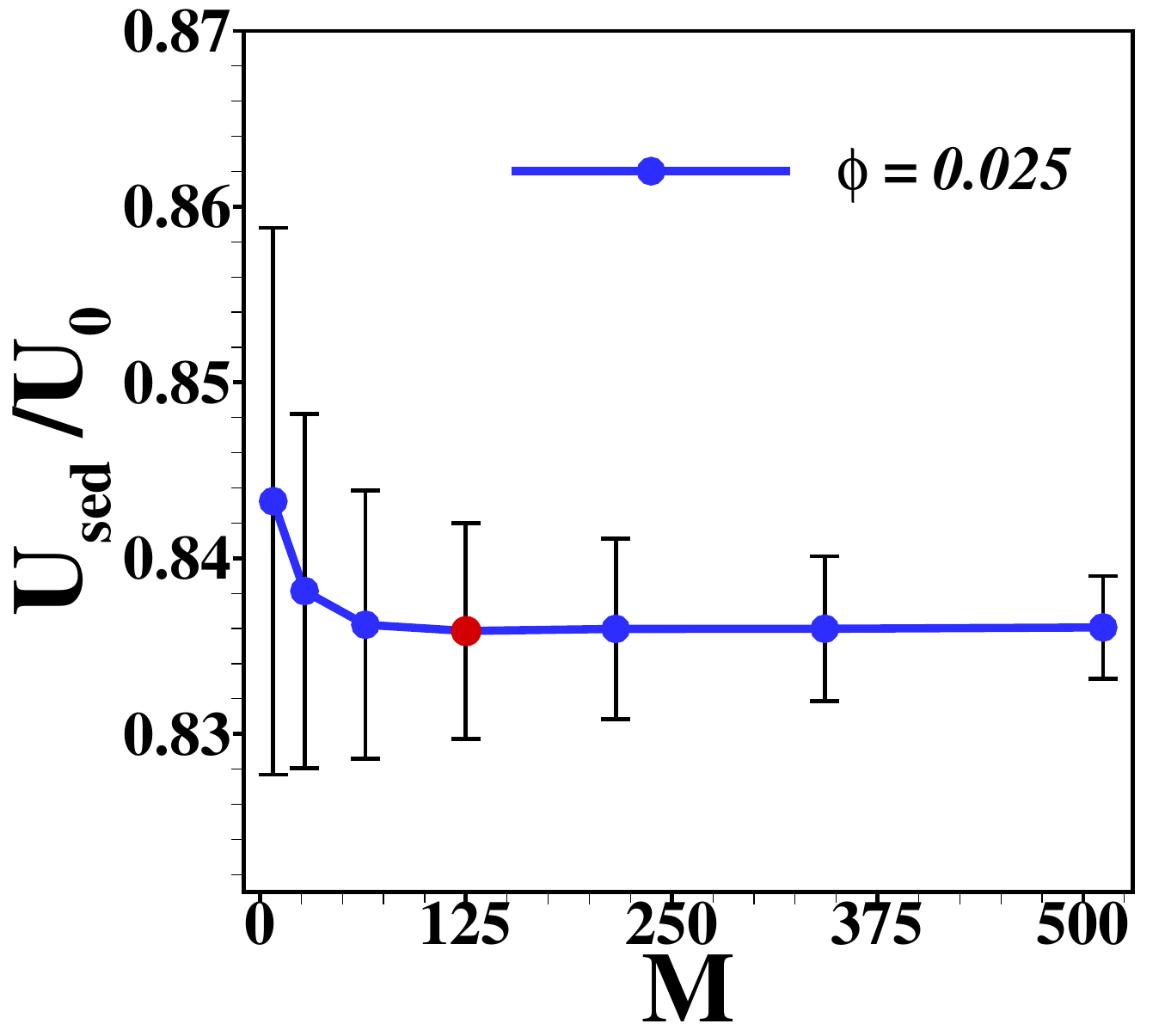}}
  \subfigure[]{\includegraphics[width=0.45\textwidth]{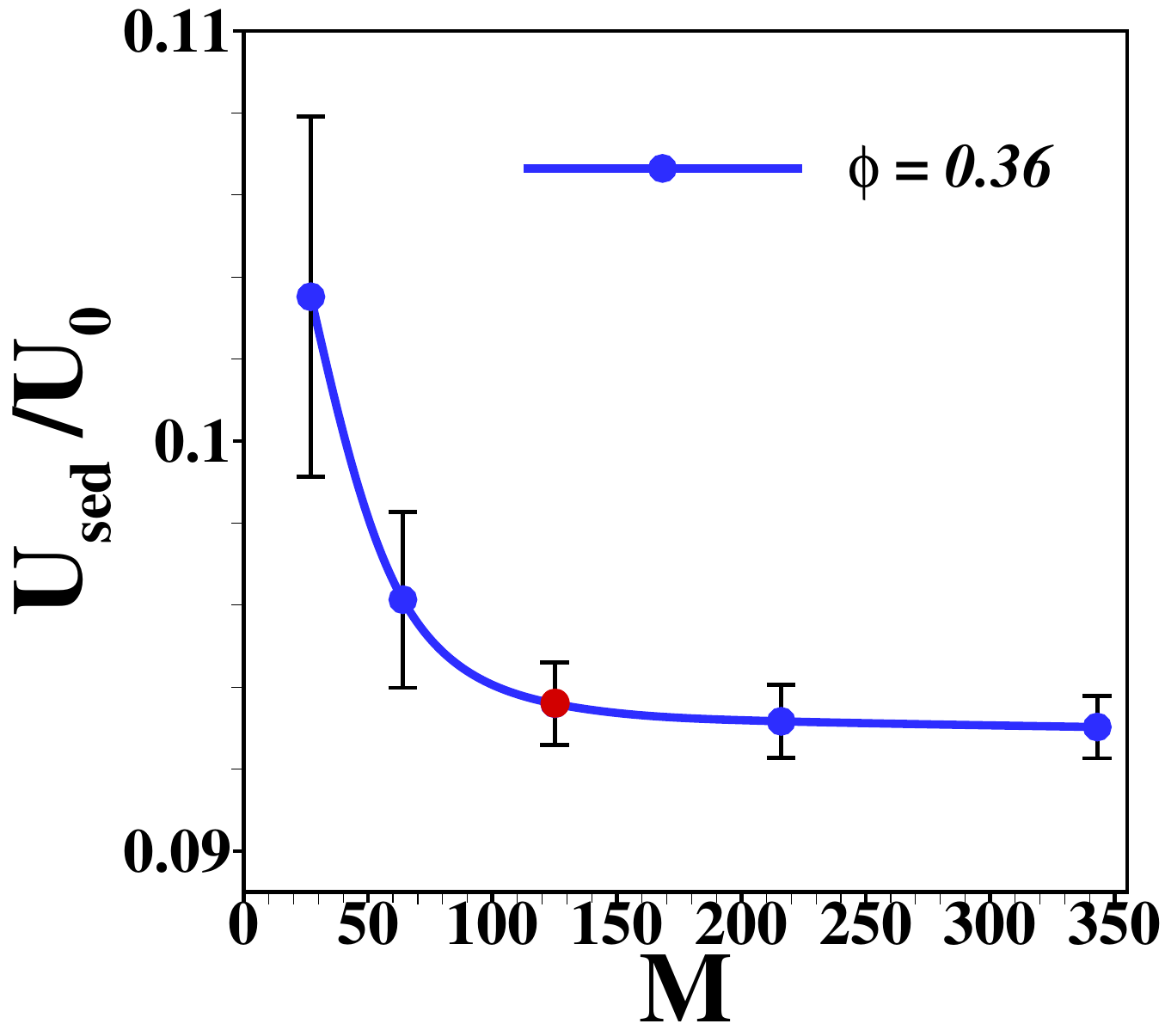}}
  \caption{Average settling velocity of spherical particles versus the number of particles $M$ 
    at two different solid volume fractions: (a) $\phi = 0.025$ and (b) $\phi = 0.36$.
    The system size $L$ is changed according with $M$ to keep the volume fraction constant,
    the settling velocity is normalized by $U_0$, and error bars represent the standard deviation of the mean settling velocity.
    The red symbols represent the value $M=125$ which we use in the rest of our simulations.
  }
  \label{domain_size}
\end{figure}
The system size determines how well the system approximates an infinite suspension and whether artificial effects distort the settling dynamics
\cite{climent2003numerical,lu2023dynamics}.
To ensure the system size is sufficiently large we study its influence on the average sedimenting velocity of spherical particles in two cases:
a dilute suspension ($\phi = 0.025$) and a moderately dense suspension ($\phi = 0.36$).
In both cases we vary the number of particles, $M$, and the system size, $L$, while keeping the volume fraction constant
and measure the average sedimenting velocity.
The results indicate that in both cases, a small box size significantly affects the mean settling velocity, see Figs.\ \ref{domain_size}a,b.
However, as the domain size increases it soon reaches the asymptotic value.
In the rest of the work we use $M=125$ so the finite sizes effects are small while the computational cost is still moderate.
The red symbols in Figs.\ \ref{domain_size}a,b indicate the selected box size used in our simulations for different solid volume fractions.

\subsection{Sedimentation velocity of monodisperse suspension of spheres and cubes}

\begin{figure}[h]
  \centering

  \subfigure[]{\includegraphics[width=0.22\textwidth]{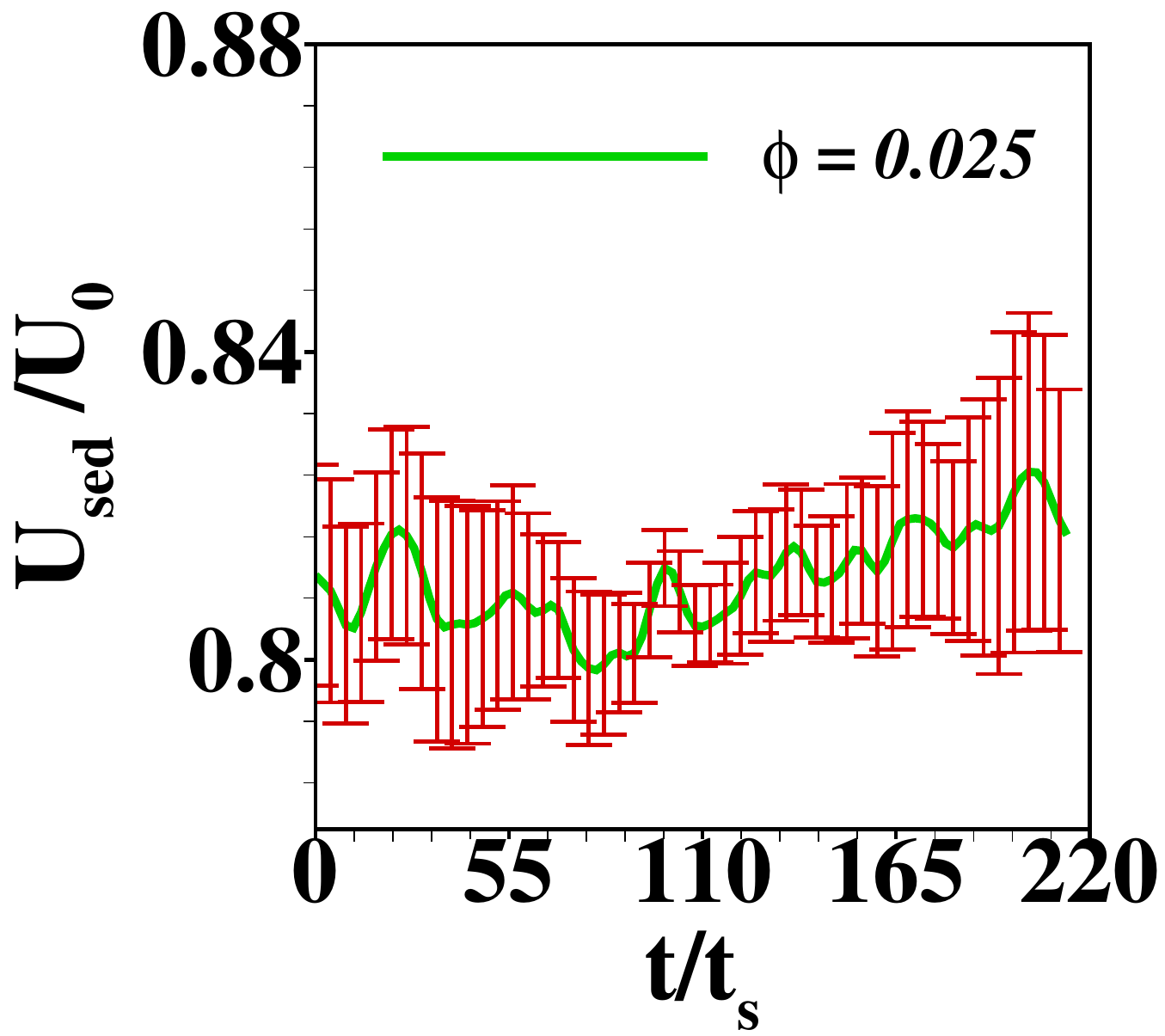}}
  \subfigure[]{\includegraphics[width=0.22\textwidth]{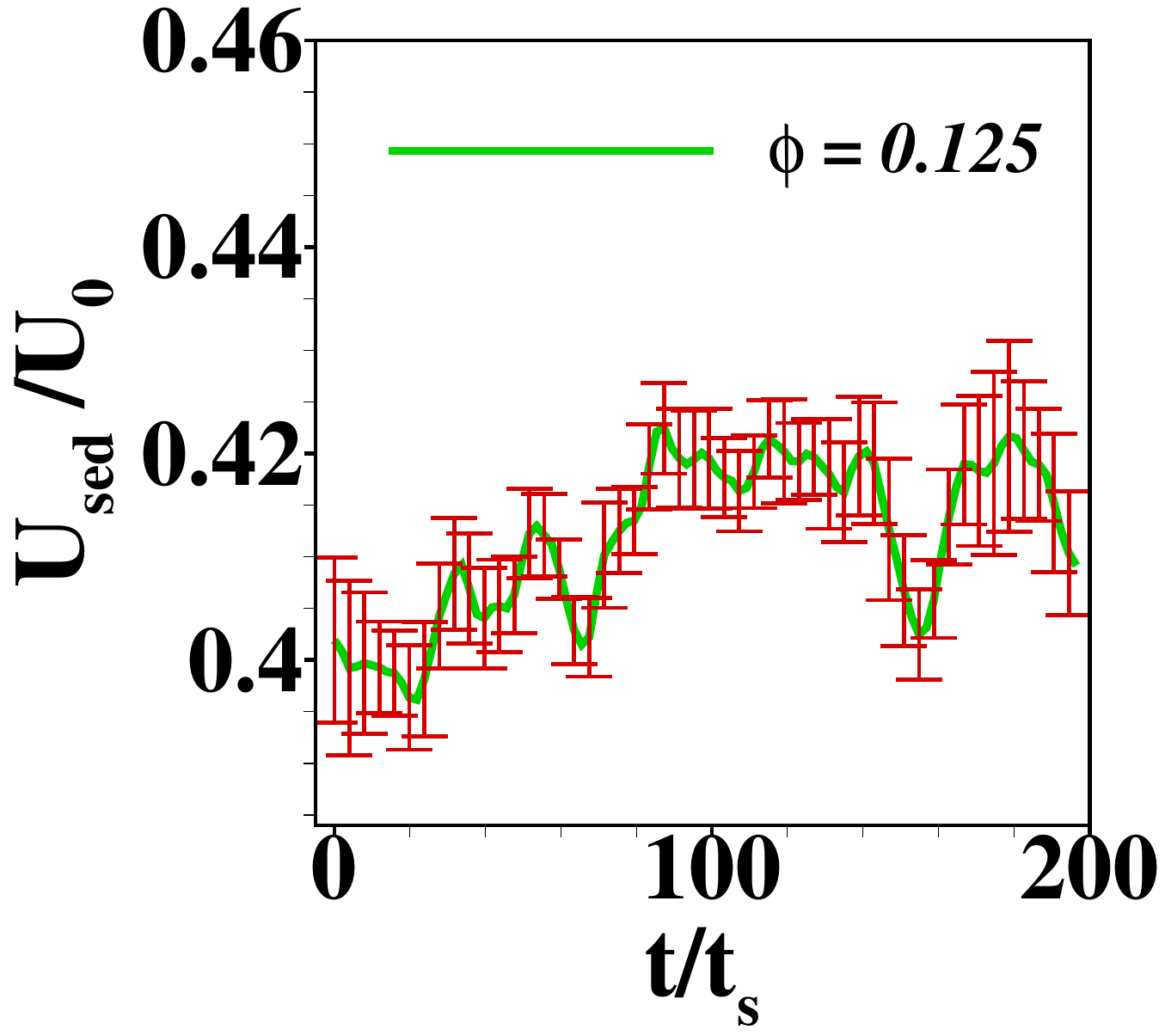}}
  \subfigure[]{\includegraphics[width=0.22\textwidth]{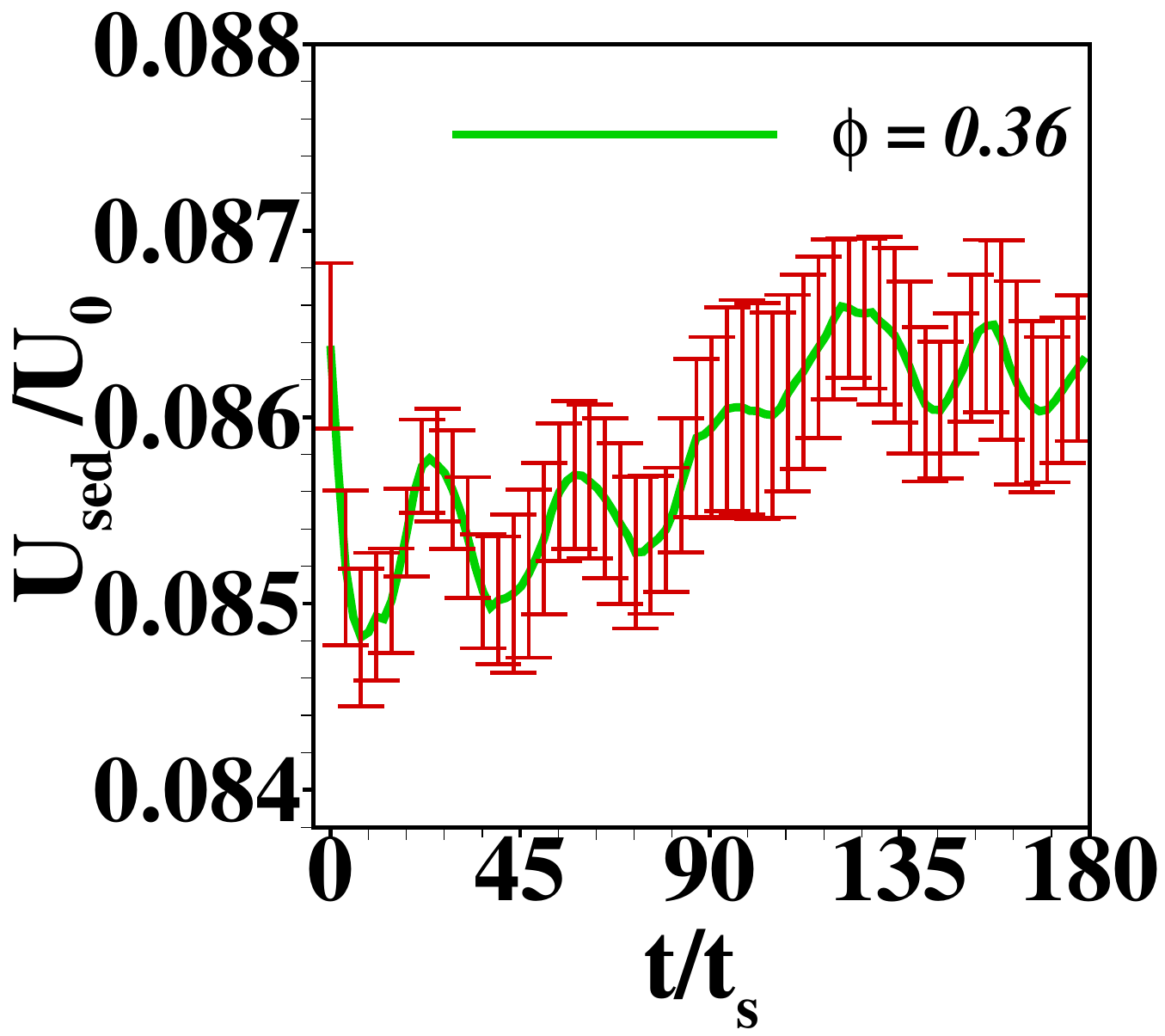}}  
  \subfigure[]{\includegraphics[width=0.22\textwidth]{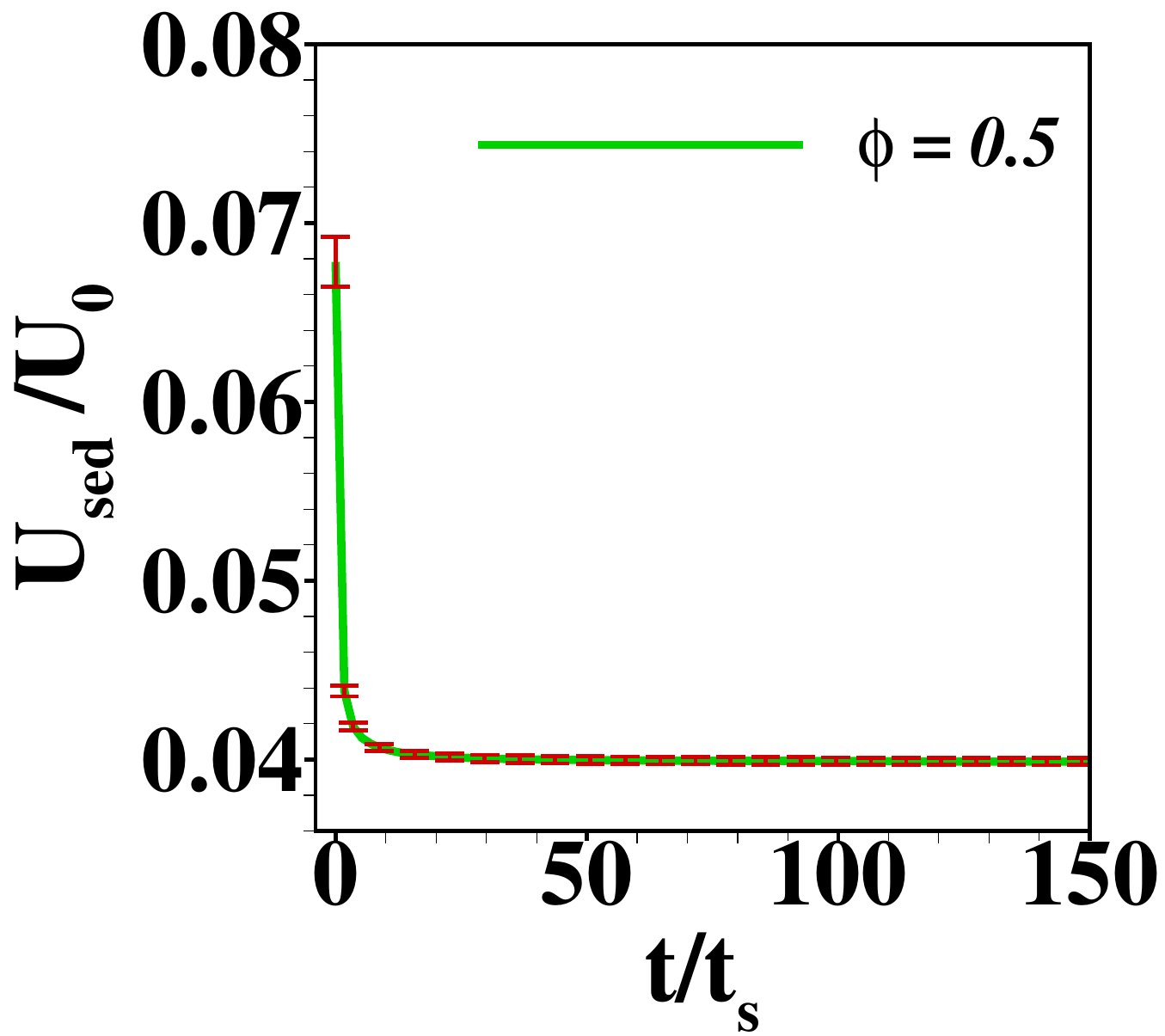}}


  \subfigure[]{\includegraphics[width=0.22\textwidth]{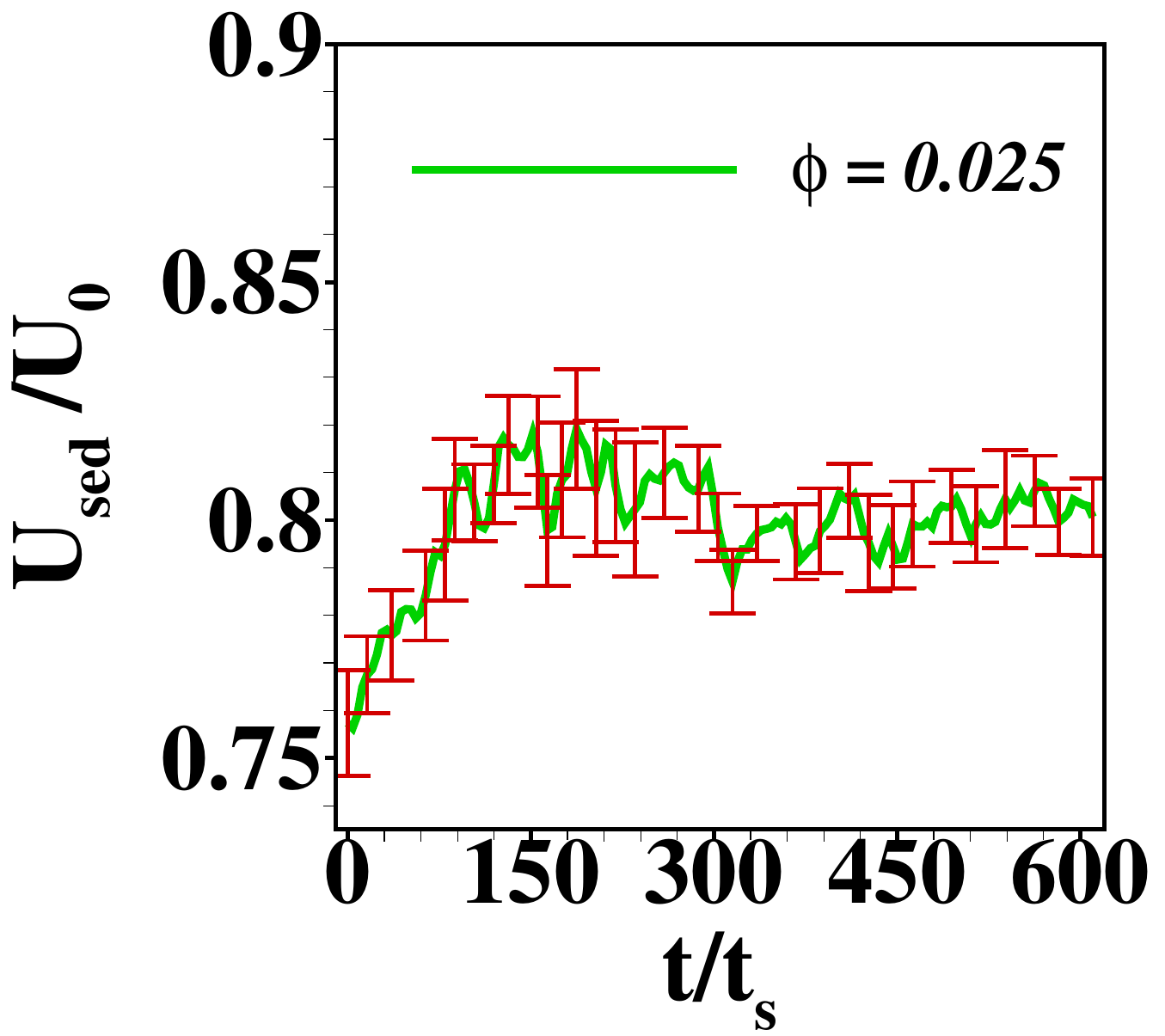}}
  \subfigure[]{\includegraphics[width=0.22\textwidth]{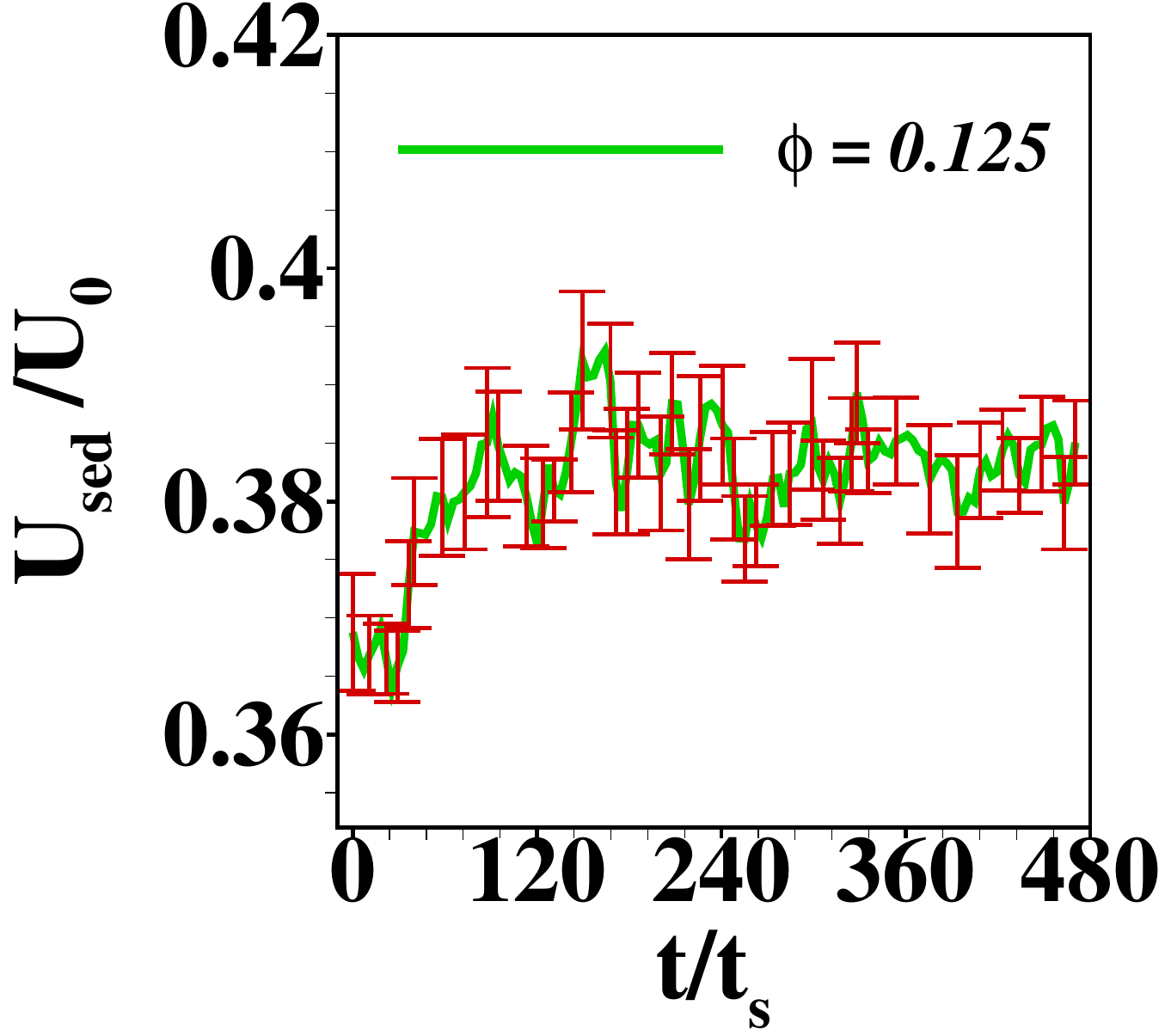}}
  \subfigure[]{\includegraphics[width=0.22\textwidth]{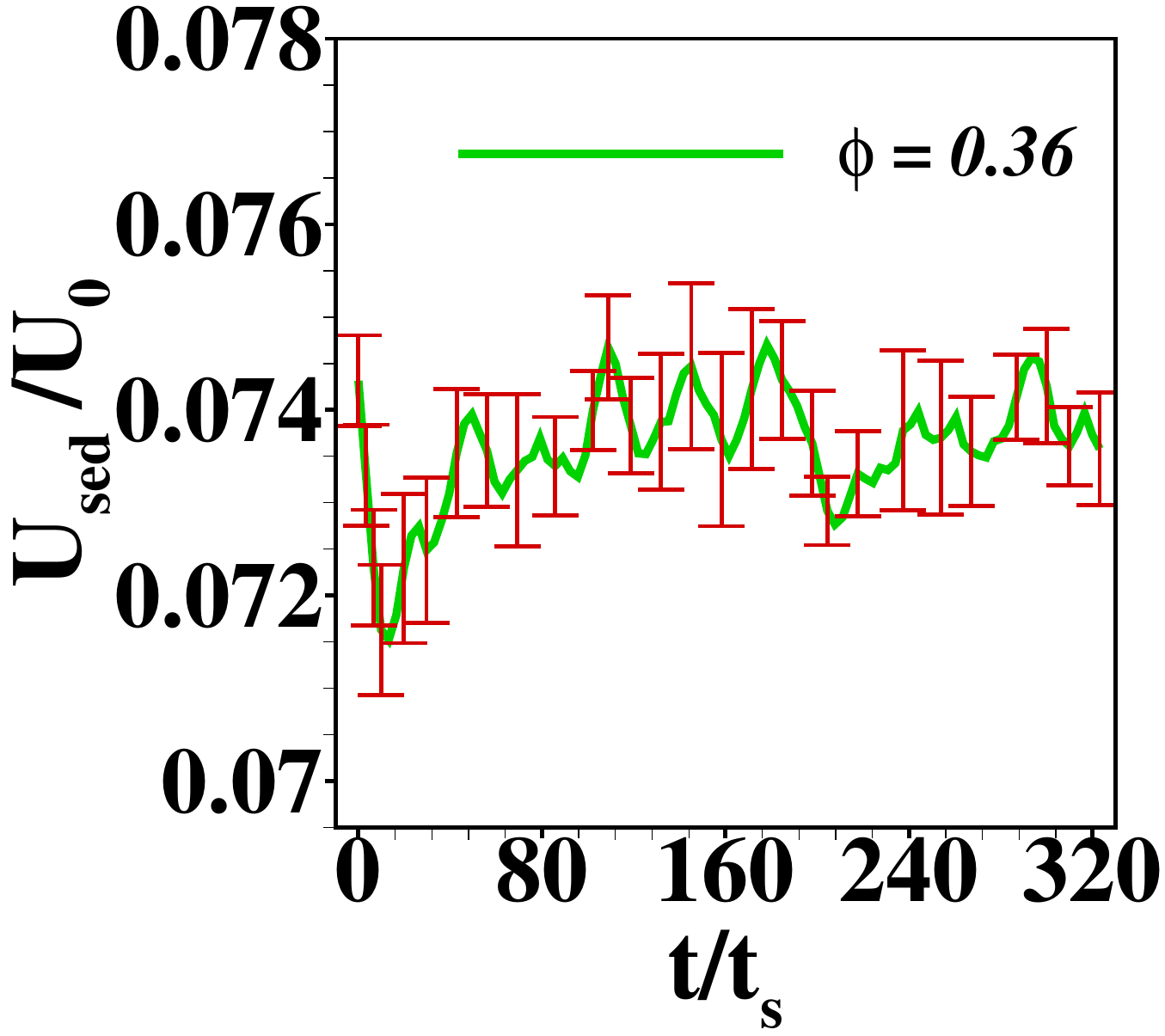}}
  \subfigure[]{\includegraphics[width=0.22\textwidth]{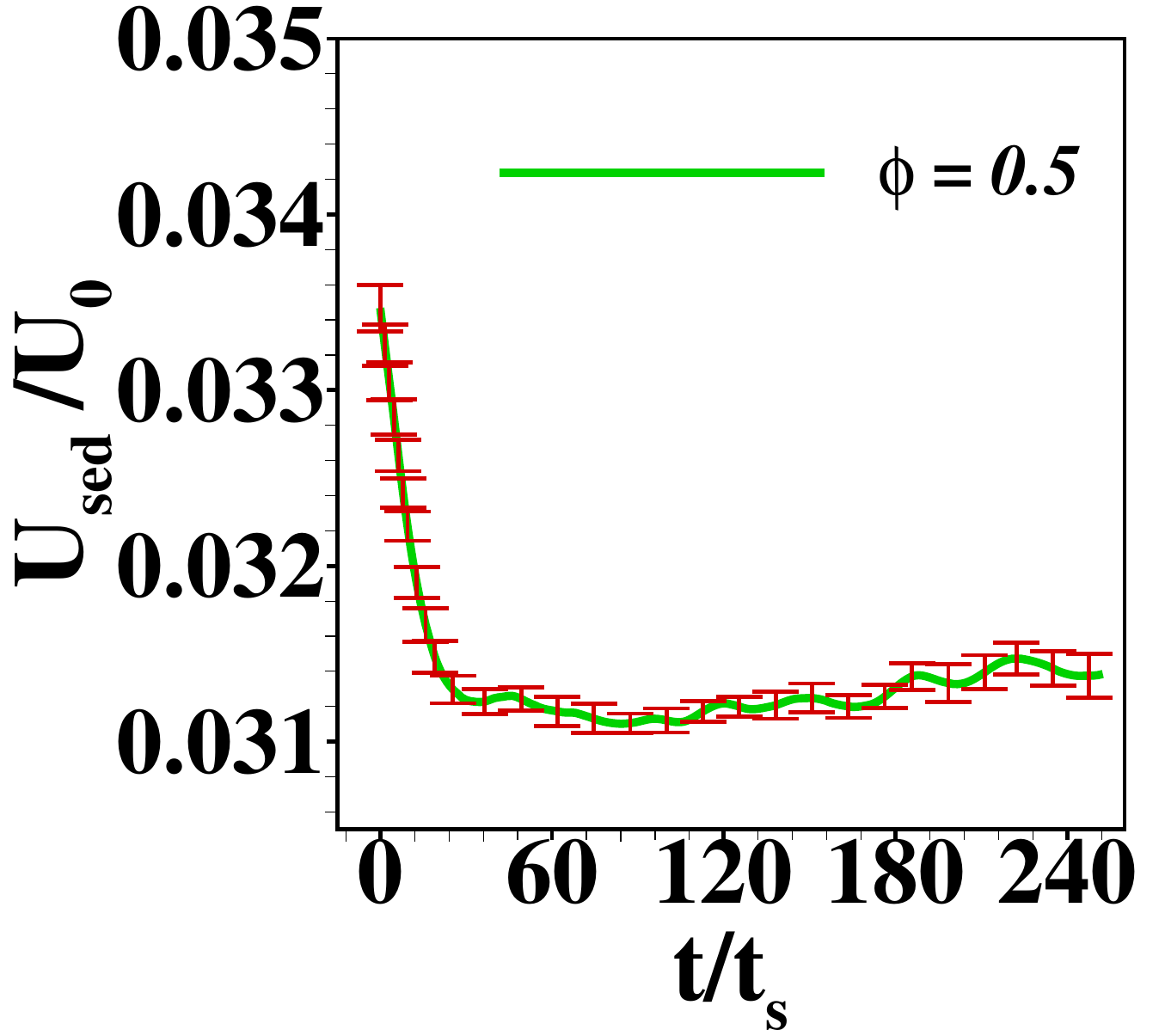}}
  \caption{Dimensionless mean sedimentation velocity of monodisperse spheres (upper row) and cubes (lower row) at various volume fractions. The error bars indicate the standard error of the mean settling velocity for several independent random configurations.}
\label{fig2}
\end{figure}
The microstructure of suspension evolves in time during sedimentation from an initial equilibrium distribution to a new steady-state distribution; therefore the sedimentation velocity,
\eqn{
  U_{\text{sed}}(t)= \Biggl \langle \frac{1}{N} \sum_{i=1}^{N} U_i(t) \Biggr \rangle,
}
ensemble averaged over the initial configurations, is time-dependent. To obtain the mean sedimentation velocity, an ensemble average (denoted by $\langle . \rangle $) of different initial configurations of the particles is considered. These configurations were randomly sampled from an equilibrium hard-particles distribution function by the Monte-Carlo method \cite{metropolis1953equation}. The initial equilibrium averages were taken over $15$ independent initial configurations. The time evolution of $U_{\text{sed}}(t)$ for randomly distributed suspension of spheres and cubes at various solid volume fractions from equilibrium to steady state, is shown in Fig.~\ref{fig2}.
The time-dependent settling velocity is normalized by the settling velocity of an isolated particle, $U_0$, (sphere/cube for corresponding cases) falling under gravity, whereas, the time is normalized with $t_s=R/U_0$.
The characteristic length scale $R$, is already mentioned in Table \ref{table:Table 1}. The settling velocity of monodisperse suspension of spheres and cubes fluctuates slightly as time progress. Fig.~\ref{fig2} shows that the mean sedimentation velocity of suspension of spheres is always slightly higher at all volume fractions compared to cubic suspension.
As reported in the study of Seyed-Ahmadi and Wachs \cite{seyed2019dynamics,seyed2021sedimentation}, anisotropic particles (i.e., cubes, tetrahedrons etc.) always produce higher rotational displacements due to their sharp edges and rapid variation of hydrodynamic loads. Thus cubic suspension causes a strong upward flux to hinder the settling process as compared to spheres. Fig.~\ref{fig2} also shows the deviation of settling velocity of random initial suspension from its mean as time progresses (indicated by error bars).

During the settling process the particles take some time in rearrange their configuration which affects the sedimentation velocity.
This relaxation time and the dispersive motion of the particles has been studied by several researchers.
The existing study of Kuusela et al.\ \cite{kuusela2005steady} demonstrated that the relaxation time in the vertical direction for a sphere suspension follow approximately a $\phi^{-1/2}$ relaxation in the full range of volume fractions. A sharp decrease in the relaxation time is also observed in the investigation of Hamid et al.\ \cite{hamid2013sedimentation} at $\phi > 0.3$ in the direction of the gravity, mainly because of the ordering of the particles, whereas a jump in relaxation time at $\phi \approx 0.5$ could occur because of very long-range ordering of the particles. At moderate range of volume fractions (i.e., $ 0.1 \leq \phi \leq 0.4$), the intensity of hydrodynamic interactions between particles increases, which, in-turn enhance the velocity fluctuations in the direction parallel and perpendicular to gravity. As a consequence, a significant time is required for the particles to relax in order to reach an equilibrium state.
With the additional increment in $\phi$, the settling process is strongly hindered and the settling particles produce a weaker upward flux of the surrounding fluid in a tightly packed space. The velocity fluctuations in both directions and hence, the relaxation time diminish drastically. As a result, both the suspensions quickly reach the steady state at $\phi \simeq 0.5$. A detailed analysis of the velocity fluctuations for both suspensions will be discussed in the next section.

\begin{figure}
  \centering
  \includegraphics[width=0.5\textwidth]{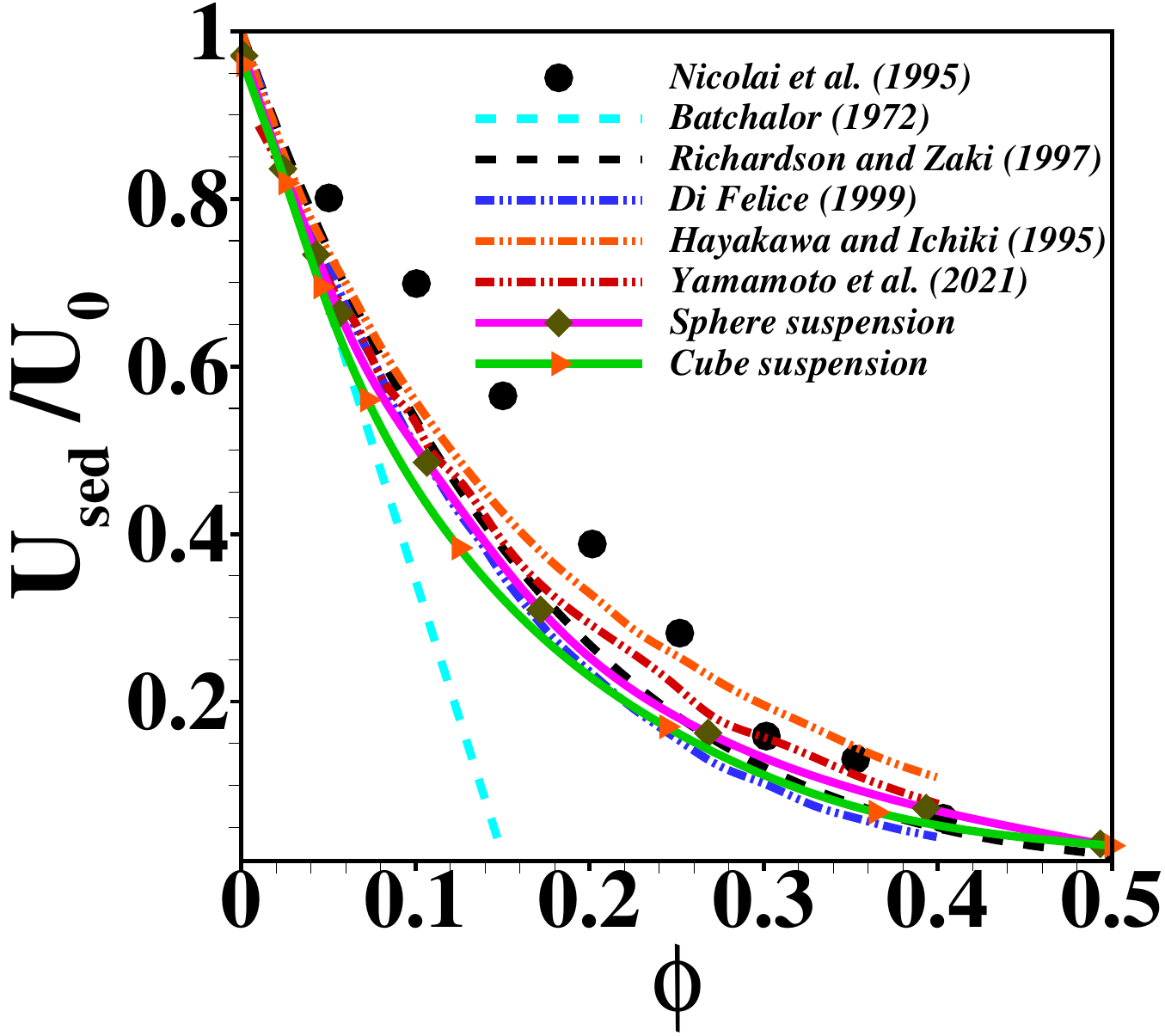}
  \caption{ Average settling velocity $U_{\text{sed}}$ of the particles normalized by the Stokes velocity $U_0$ as a function of volume fraction ($\phi$) to highlight the hindered settling of the particles. Solid lines denote our simulation result; dashed lines show various theoretical predictions of Hayakawa and Ichiki \cite{hayakawa1995statistical}, Richardson and Zaki \cite{richardson1997sedimentation}, Di Felice \cite{di1999sedimentation}, Batchelor \cite{batchelor1972sedimentation}, simulation result of Yamamoto et al.\ \cite{yamamoto2021smoothed}, and the symbols represent the experimental data of Nicolai et al.\ \cite{nicolai1995particle}. Here, the error bars are smaller than the symbols' size.}
  \label{fig1}
\end{figure}

In Fig.\ \ref{fig1}, we present the normalized average settling velocity $U_\text{sed}/U_0$, as a function of the solid volume fraction, where $U_{\text{sed}}$ and $U_0$ denote the ensemble average of the suspension settling velocity and the terminal settling velocity of an isolated sphere, respectively.
We also compare our computational result for spheres with the existing theoretical predictions \cite{hayakawa1995statistical,richardson1997sedimentation,di1999sedimentation,batchelor1972sedimentation}, experimental study \cite{nicolai1995particle} and simulation data \cite{yamamoto2021smoothed,hamid2013sedimentation}, which involves many-body hydrodynamic interactions. 
We observe that the sedimentation velocity decreases with increasing volume fraction.
Since both the particles and the fluid are incompressible as particles sediment an upward flow to balance the flow of mass
is generated which hinders the sedimentation velocity.
This effect of particle volume fraction on the average sedimentation velocity was first introduced by Batchelor \cite{batchelor1972sedimentation}.
The theoretical prediction of Batchelor assumes a uniform distribution in the separation of pairs of spheres, ignoring the fluid back flow effects, given by $U_{\text{sed}}/U_0=1-6.55 \phi$, which is only valid for low volume fractions, where long range hydrodynamic interactions can be neglected. We find a good accuracy in the sedimentation velocity for spheres with Batchelor's correlation in the range $\phi \leq 0.04$. For larger volume fractions, our simulation result overestimates the Batchelor’s model. Hayakawa and Ichiki \cite{hayakawa1995statistical} also proposed the normalized sedimentation rate in the context of a resistance problem, given by, $ \frac{U_{\text{sed}}}{U_0} = \frac{(1-\phi)^3}{ 1+2 \phi + 1.492 \phi (1-\phi)^3 } $. Our current result also agrees well with the Hayakawa-Ichiki model for $\phi \leq 0.05$. The simplest semi-empirical relation for hindered settling function has been proposed by Richardson and Zaki \cite{richardson1997sedimentation} as, $U_{\text{sed}}/U_0=(1- \phi)^n$, which gives good predictions at higher $\phi$. Several studies have found different exponents ranging from $n=4.7$ to $6.55$ for the Stokes flow, whereas in our study the best suited exponent is $n=5.9$ for spheres and $n=6.2$ for cubes. Finally, we observe that the cubes always sediment slower than the sphere suspensions for all volume fractions, except possibly at very dilute regime.


\subsection{Velocity fluctuations}
\label{Vel_fluctuation}


\begin{figure}[h]
  \centering
  \includegraphics[width=0.31\textwidth]{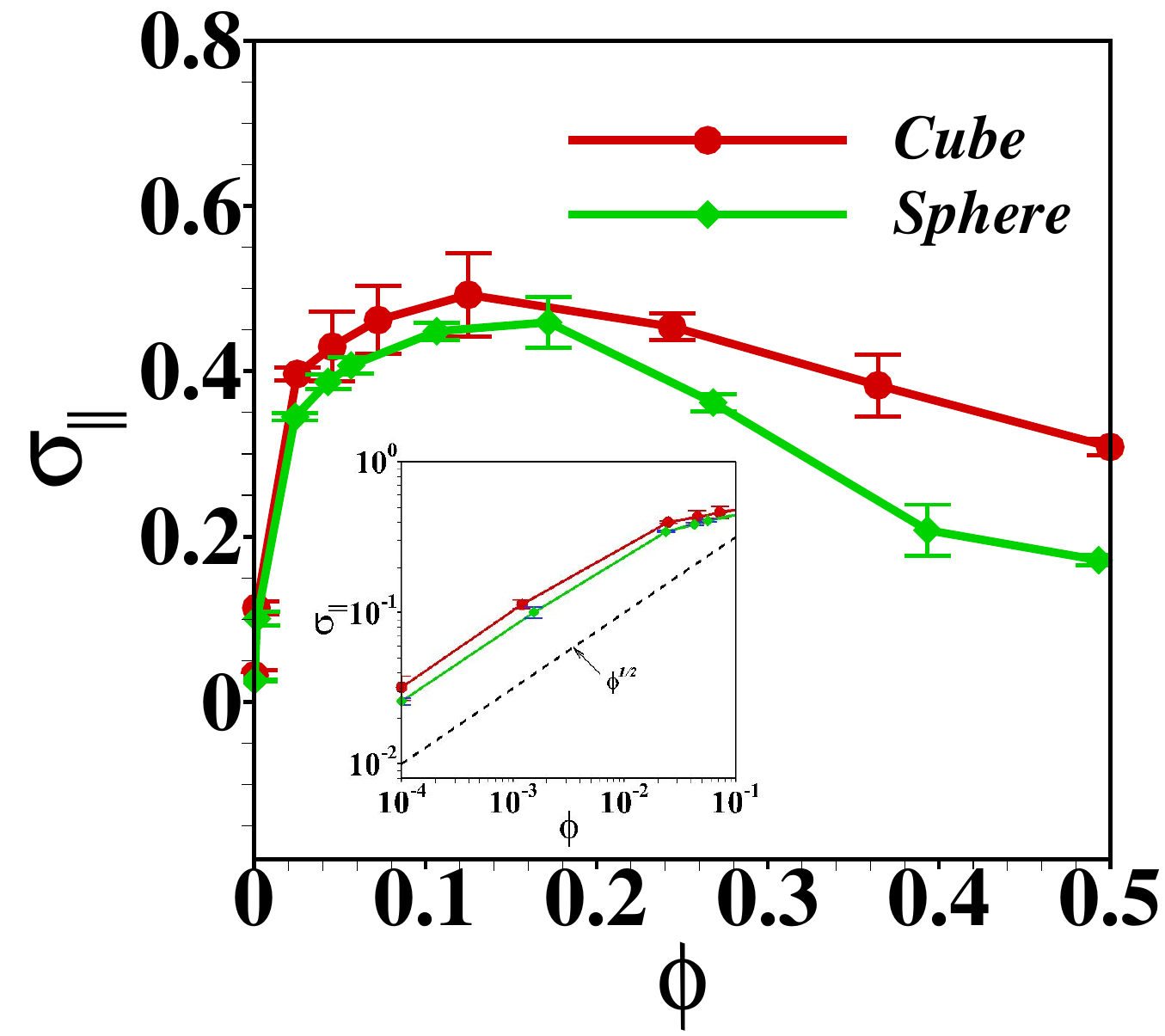}
  \includegraphics[width=0.31\textwidth]{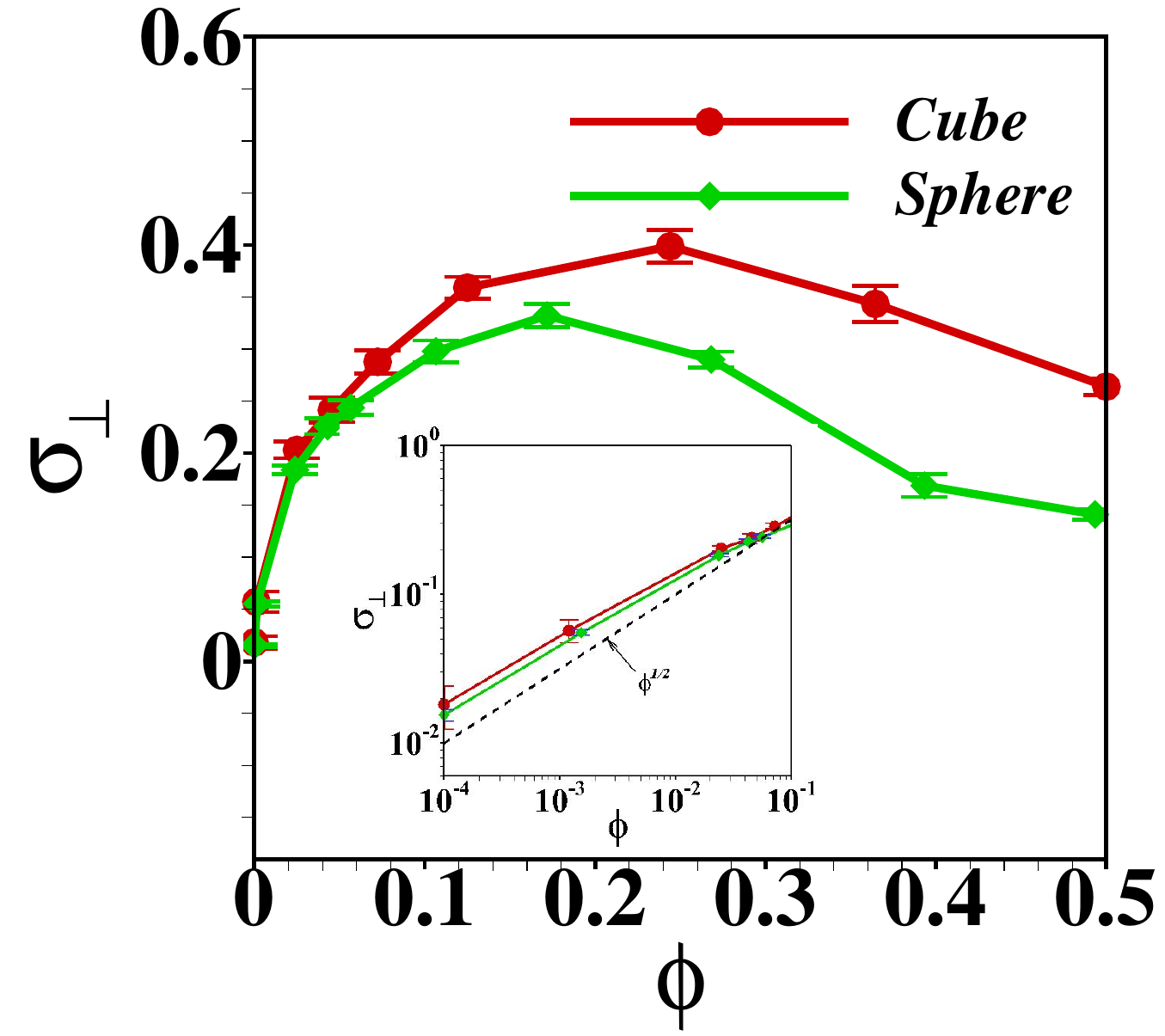}
  \includegraphics[width=0.31\textwidth]{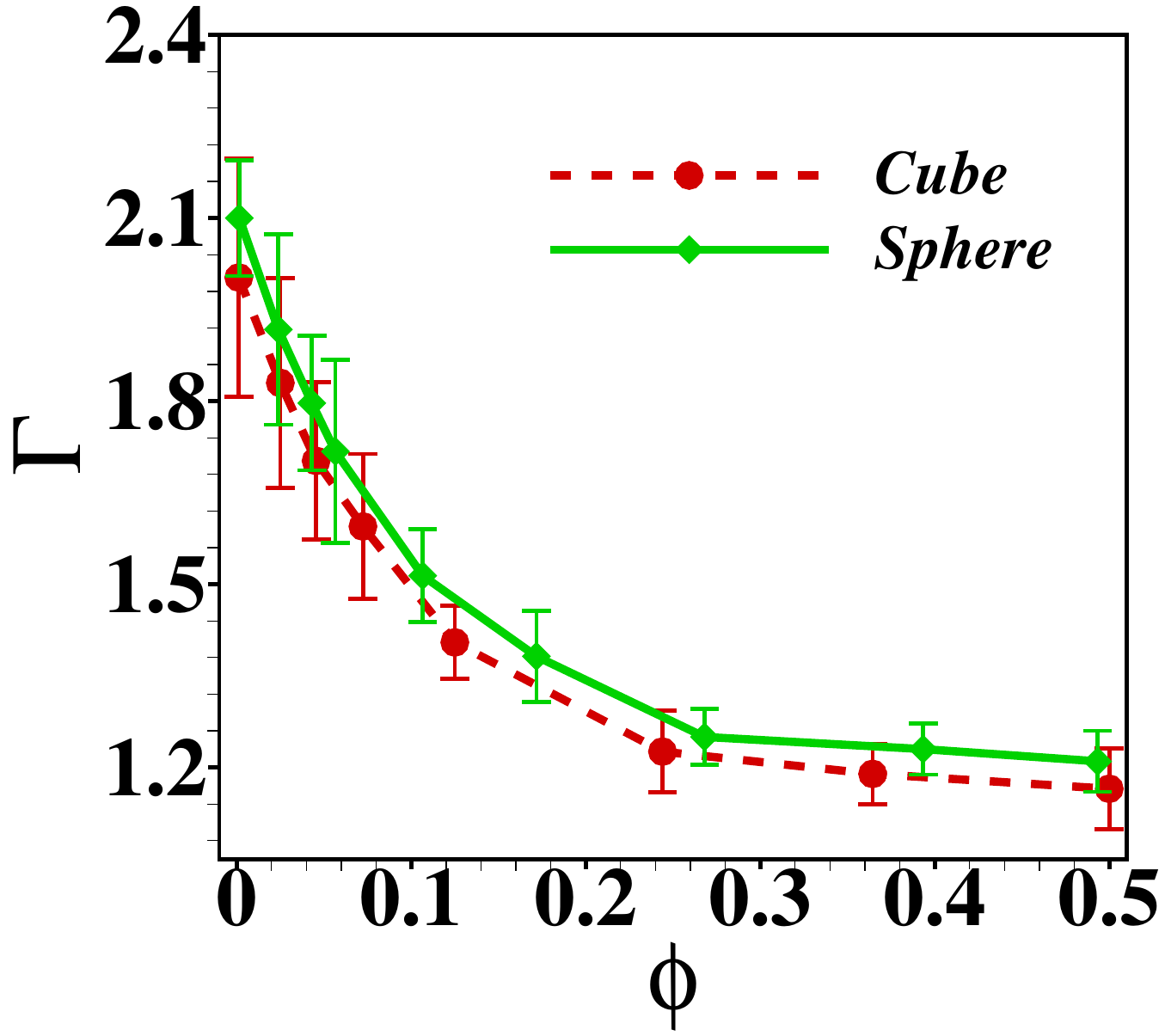}
  \caption{Dimensionless hydrodynamic velocity fluctuations of particles as a function of volume fraction ($\phi$) are shown for (a) the settling direction ($\sigma_{\parallel}$) and (b) the direction perpendicular to settling ($\sigma_{\perp}$). The inset illustrates the dependence of velocity fluctuations on $\phi^{1/2}$ for $\phi \leq 0.1$. (c) The anisotropy in velocity fluctuations, $\Gamma = \sigma_{\parallel} / \sigma_{\perp}$, is plotted as a function of volume fraction ($\phi$). }
\label{velfluc}  
\end{figure}
During settling, particle velocities fluctuate both vertically and horizontally due to inter-particle hydrodynamic interactions. Vertical velocity fluctuations reflect temporal deviations from the mean sedimentation velocity, while horizontal fluctuations, occurring perpendicular to gravity, are characterized by the mean velocities in the $XY$ plane. The time-averaged standard deviations of particle velocities parallel and perpendicular to gravity are quantified as $\Delta U_z = \sqrt{ \langle [U_{iz} - U_{\text{sed}}]^2 \rangle }$ and $\Delta U_{xy} = \sqrt{ \langle U_{ix}^2 + U_{iy}^2 \rangle }$, respectively. Here, $U_{iz}$ represents the temporal velocity of the $i$-th particle along the gravitational direction, while $U_{ix}$ and $U_{iy}$ are the velocities in the horizontal directions. These standard deviation values are normalized by the corresponding settling velocity of each suspension, commonly referred to as 'relative fluctuation' \cite{climent2003numerical,nicolai1995particle}. Figures \ref{velfluc}a and \ref{velfluc}b display dimensionless velocity fluctuations of particles in the settling direction ($\sigma_{\parallel} = \Delta U_z / U_{\text{sed}}$) and in the perpendicular direction ($\sigma_{\perp} = \Delta U_{xy} / U_{\text{sed}}$), respectively. The fluctuations exhibit similar qualitative behavior in both directions. At low volume fractions ($\phi < 0.04$), $\sigma_{\parallel}$ and $\sigma_{\perp}$ display characteristics of the Stokes regime, with a $\phi^{1/2}$ dependence (see insets of Figs.~\ref{velfluc}a,b).
As the solid volume fraction increases, velocity fluctuations rise gradually up to a moderate range ($0.04 \leq \phi \leq 0.2$).
Beyond this range ($\phi > 0.2$), a gradual decline in fluctuations is observed for both directions.
Additionally, velocity fluctuations are slightly higher for cubes compared to spheres, with the differences between the two shapes becoming more pronounced as $\phi$ increases.
As the volume fraction increases within the range ($0.04 \leq \phi \leq 0.2$),
particles are more crowded and collision events, where steric interactions push the particles apart, become more common.
Consequently, momentum is significantly transferred from the axial direction to the transverse direction, which amplifies velocity fluctuations in the transverse direction \cite{yao2021effects}.
Notably, even a slight reorientation of a cube can generate considerable hydrodynamic torques, leading to rotations that further enhance this effect \cite{seyed2019dynamics}.
As a consequence, cube seems to experience higher transverse velocity fluctuations compared to spheres, thus facilitating a more efficient transfer of momentum from the axial to the transverse direction.
Further increment in $\phi$, the average inter-particle distance, which is inversely proportional to ($\phi$), diminishes, causing particles to become more densely packed \cite{zaidi2018particle}.
Beyond ($\phi > 0.2$), this dense packing restricts particle motion, resulting in a reduction in velocity fluctuations.
In this range, the reduced inter-particle spacing disrupts the fluid structure required for particle clustering, while the frequency of hydrodynamic collisions rises, further suppressing velocity fluctuations.

The increase in velocity fluctuations with rising solid volume fraction is closely associated with the degree of anisotropy in these fluctuations for both cubic and spherical suspensions.
This anisotropy, defined as the ratio of vertical to horizontal velocity fluctuations (\(\Gamma = \sigma_{\parallel} / \sigma_{\perp}\)),
is shown in Fig.~\ref{velfluc}c.
In highly dilute conditions, anisotropy is notably high for both particle shapes, primarily due to restricted lateral movement in this regime. For dilute suspensions (\(\phi \leq 0.05\)), \(\Gamma\) ranges between 1.7 and 2.1 for both cases, aligning with the experimental findings of Nicolai and Guazzelli \cite{nicolai1995effect} on non-Brownian spheres under very dilute conditions. As the solid volume fraction increases to moderate levels (\(\phi \approx 0.2\)), velocity fluctuations intensify in both vertical and horizontal directions, leading to a substantial reduction in anisotropy.
In this range, the anisotropy decreases 
compared to the dilute regime, with $\Gamma$ approaching approximately $1.3$ for cubic suspensions.
A similar trend was also observed in the studies of Nicolai and Guazzelli \cite{nicolai1995effect} and Ladd \cite{ladd1996hydrodynamic}.
In a sedimentation study at finite Reynolds numbers by Seyed-Ahmadi and Wachs \cite{seyed2021sedimentation} was observed
that the higher rotational velocities of cubes contributed to the transfer of momentum and energy from the gravity to the transverse direction,
which led to a lower velocity fluctuations anisotropy in cubic suspensions than in spherical suspensions.
Here, we observe the same effect in zero Reynolds number regime.
As $\phi$ increases further to $0.5$, the anisotropy continues to decline due to more frequent and multidirectional inter-particle collisions.

\subsection{Probability density function of particle sedimentation velocity}
\label{PDF1}

\begin{figure}[h]
  \centering
  \subfigure[]{\includegraphics[width=0.31\textwidth]{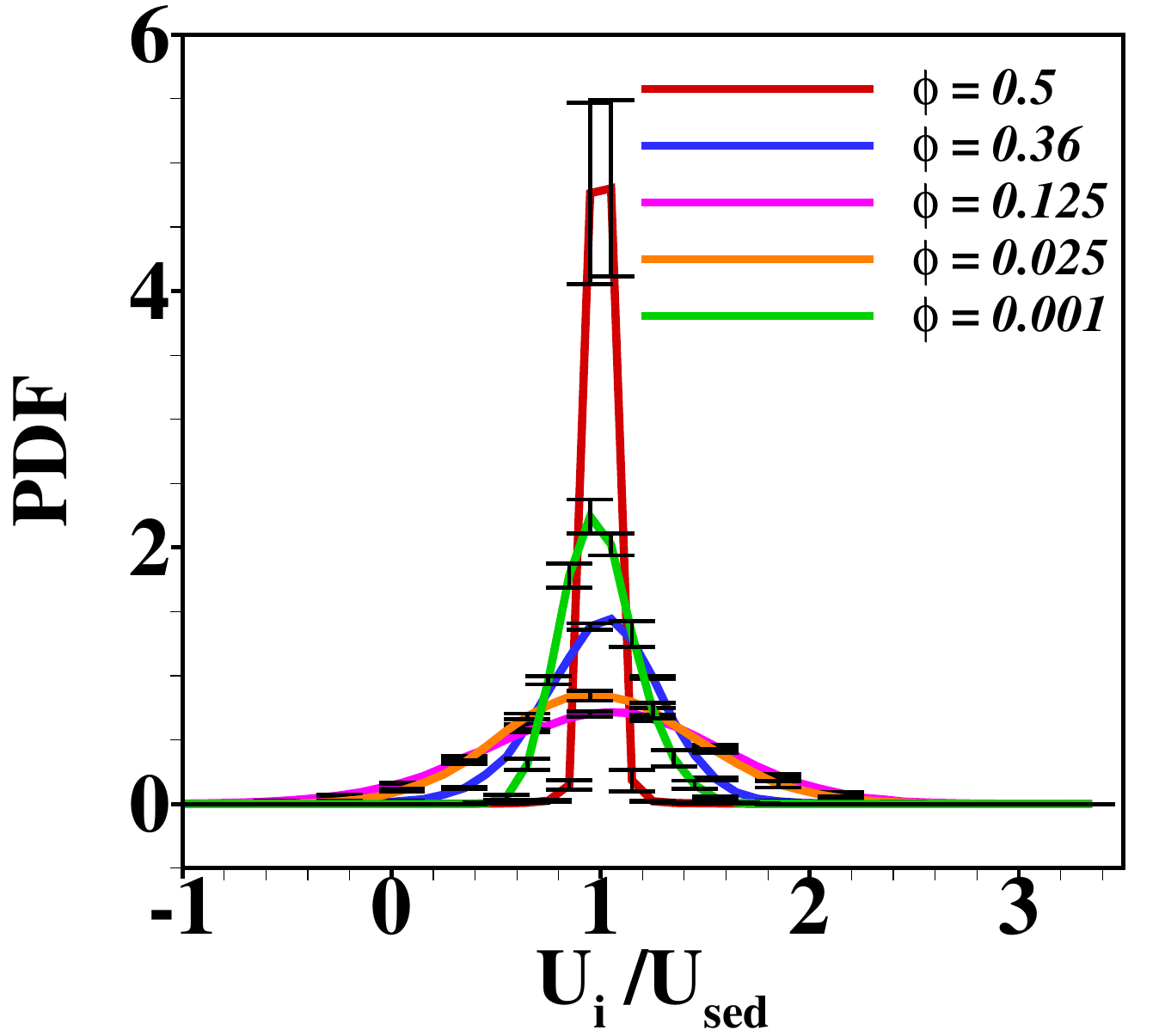}}
  \subfigure[]{\includegraphics[width=0.31\textwidth]{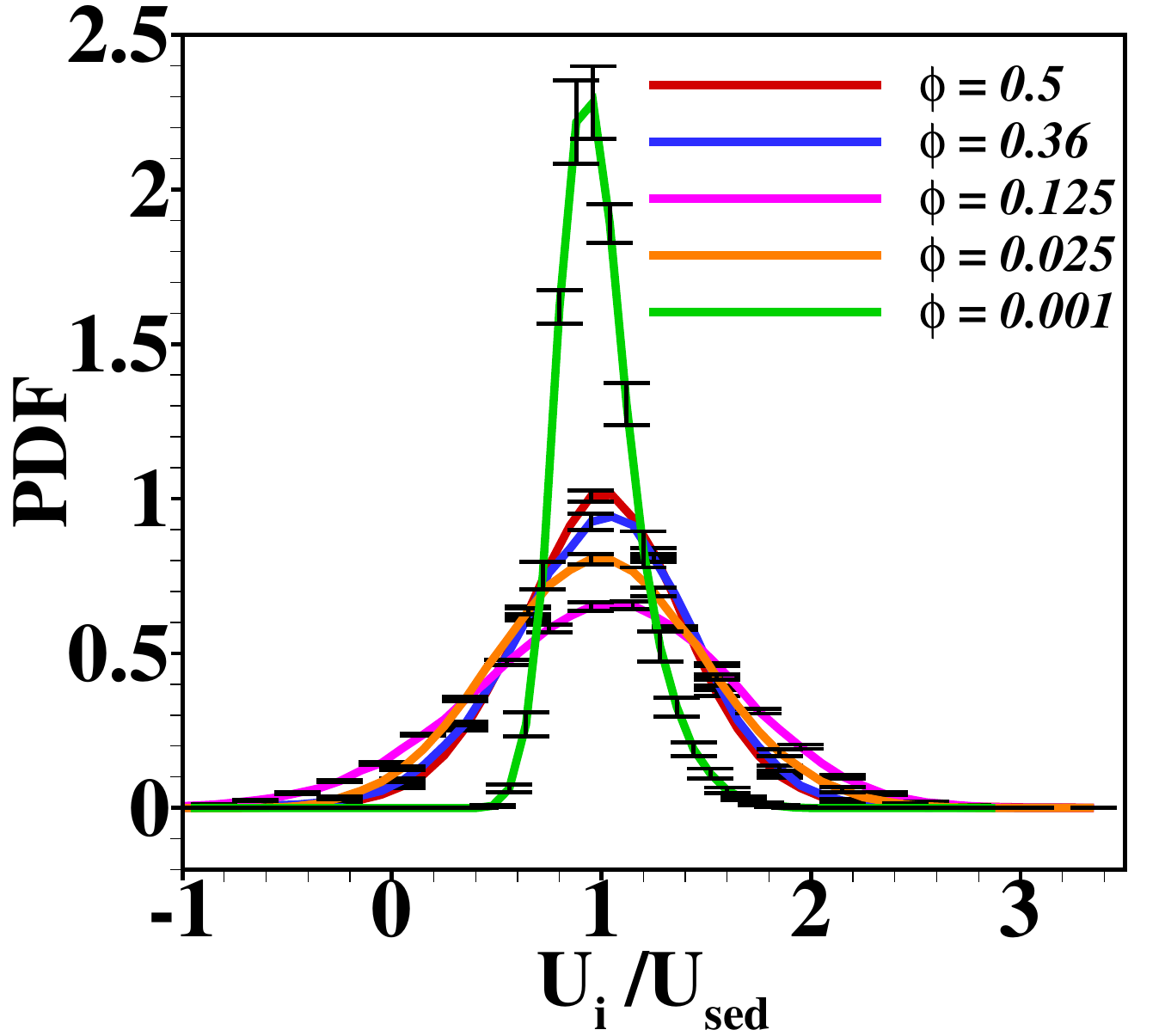}}
  \subfigure[]{\includegraphics[width=0.31\textwidth]{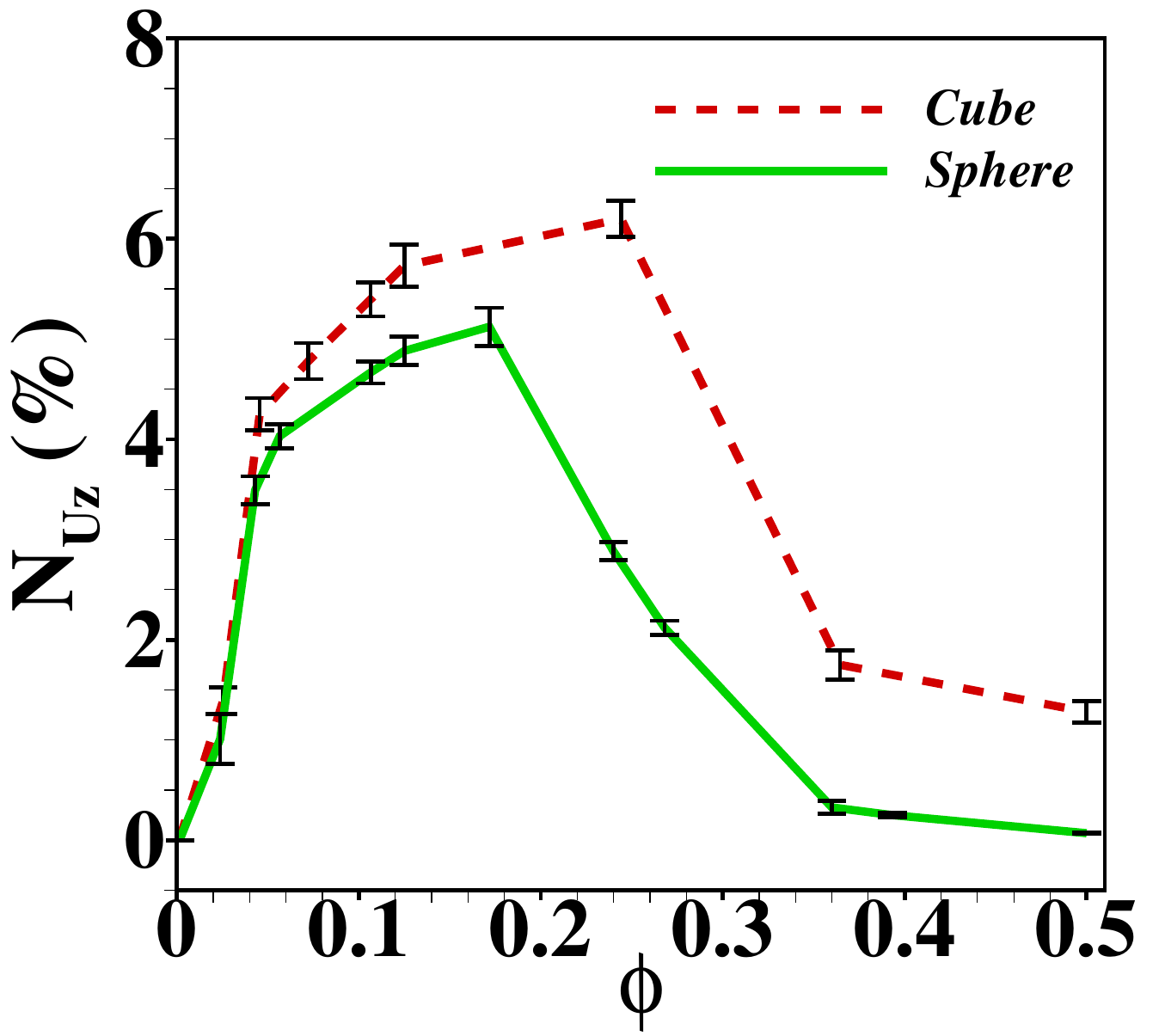}}
  \caption{The probability density function (PDF) of vertical velocity is presented for suspensions of (a) spherical and (b) cubic particles across different solid volume fractions. Negative values of $U_i/U_{\text{sed}}$ indicate that particles are moving vertically upward, against the direction of gravity. Additionally, (c) illustrates the percentage of particles moving opposite to the gravity at any given moment as a function of solid volume fraction. }
\label{pdf}
\end{figure}

To better understand the differences in velocity fluctuations between spherical and cubic particles, we investigate the instantaneous particle velocity during sedimentation.
Figs \ref{pdf}a and \ref{pdf}b illustrate the normalized probability density function (PDF) of particle velocity as a function of bulk concentration.
Notably, $U_i/U_{\text{sed}} > 1$ indicates that particles are falling faster than the average sedimentation velocity.
Positive values of $U_i/U_{\text{sed}}$ imply downward movement (i.e.\ in the direction of gravity), while negative values represent upward movement.
Note that as the particles settle, the surrounding fluid generates an upward jet due to the conservation of mass, that can drag some particles with it.
Interesting, the variance of the velocity in the vertical direction is a non-monotonous function of the volume fraction.
In a very dilute regime ($\phi \approx 0.001$), both the spherical and cubic suspensions exhibit a sharp peak, indicating that the velocity variance is small.
In this range, the larger inter-particle separation distance results in weaker hydrodynamic interactions, thus deviation from the single particle dynamics are small.
Additionally, the percentage of upward-moving particles ($N_{U_z}$) is calculated by integrating the area under the PDF of the settling velocity.
Figure \ref{pdf}c illustrates that, in the highly dilute regime, the fraction of upward-moving particles approaches to nearly zero for both types of particles.

As the volume fraction increases, velocity fluctuations increase, causing particles to disperse more and collide more frequently. At moderate concentrations, hydrodynamic interactions among nearby settling particles and inter-particle collisions create a strong upward jet, which can lift some particles (both spheres and cubes) along with it. The frequent hydrodynamic interactions and particle rebounds after collisions may also lead to individual particles moving upward.
A closer examination of the PDF of vertical particle velocity reveals that the tails become stretched as $\phi$ increases from $0.025$ to $ 0.125$, with the left tail slightly longer than the right.
Therefore, the fraction of upward-moving particles increases with rising bulk concentration in the low to moderate range.
Results in Fig.\ \ref{pdf}c indicate that in cubic suspensions, approximately
$5.3 \pm 0.2 \%$ of particles move against the settling direction, compared to only $4.6 \pm 0.1 \%$ for spherical particles at $\phi \approx 0.1$.
In this range, a closer inspection in Figs.\ \ref{pdf}a,b reveals that the peak of the PDF is slightly higher for spherical particles, indicating that a greater number of them attain velocities close to the mean sedimentation velocity.
In contrast, cubic particles exhibit a slightly stretched left tail in the negative region, showing a broader distribution of lower velocities,
see also Supplementary Figure 1.
The sharp edges and rapid changes in hydrodynamic forces due to variations in orientation make cubes more susceptible to strong translational and rotational displacements,
as well as velocity fluctuations.
Additionally, cubes tend to acquire slightly higher angular velocities than spheres, resulting in significant rotation-induced forces perpendicular to their direction of movement.
Consequently, cubic suspensions generate larger rotation-induced forces due to the higher magnitude of angular rotation, resulting in a greater fraction of particles being entrained by the upward jet compared to spherical suspensions.
At moderate $\phi$, this effect becomes more pronounced, with $6.2 \pm 0.2 \%$ of cubic particles moving upward, compared to only $2.86 \pm 0.08 \%$ for spherical particles at
$\phi \approx 0.24$ (Fig.\ \ref{pdf}c).

However, as $\phi$ increases further, the system becomes more densely packed, suppressing velocity fluctuations in both directions. Consequently, the tails of the PDF begin to contract for both particle types (Fig.\ \ref{pdf}a,b), leading to a gradual decrease in the percentage of upward-moving particles, as shown in Fig.\ \ref{pdf}c.
In the very dense regime ($\phi = 0.5$), only a small fraction of spherical particles ($0.070 \pm 0.003 \%$) exhibit instantaneous upward movement,
whereas a slightly larger fraction of cubic particles ($1.3 \pm 0.1 \%$) continue to move upward.

\subsection{Suspension microstructure}
\label{sec:microstructure}
\begin{figure}[h]
  \centering
  \subfigure[]{\includegraphics[width=0.45\textwidth]{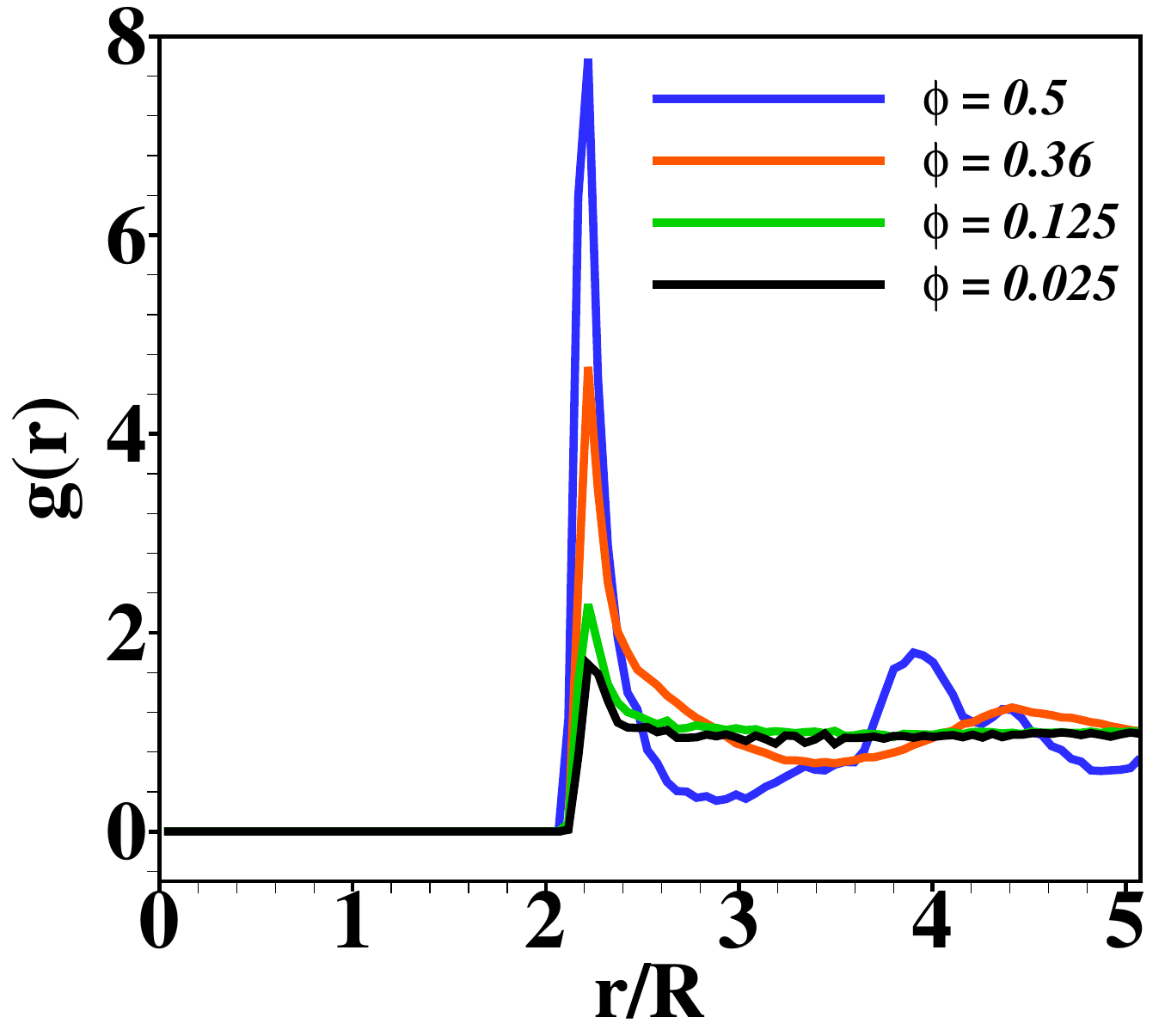}}
  \subfigure[]{\includegraphics[width=0.45\textwidth]{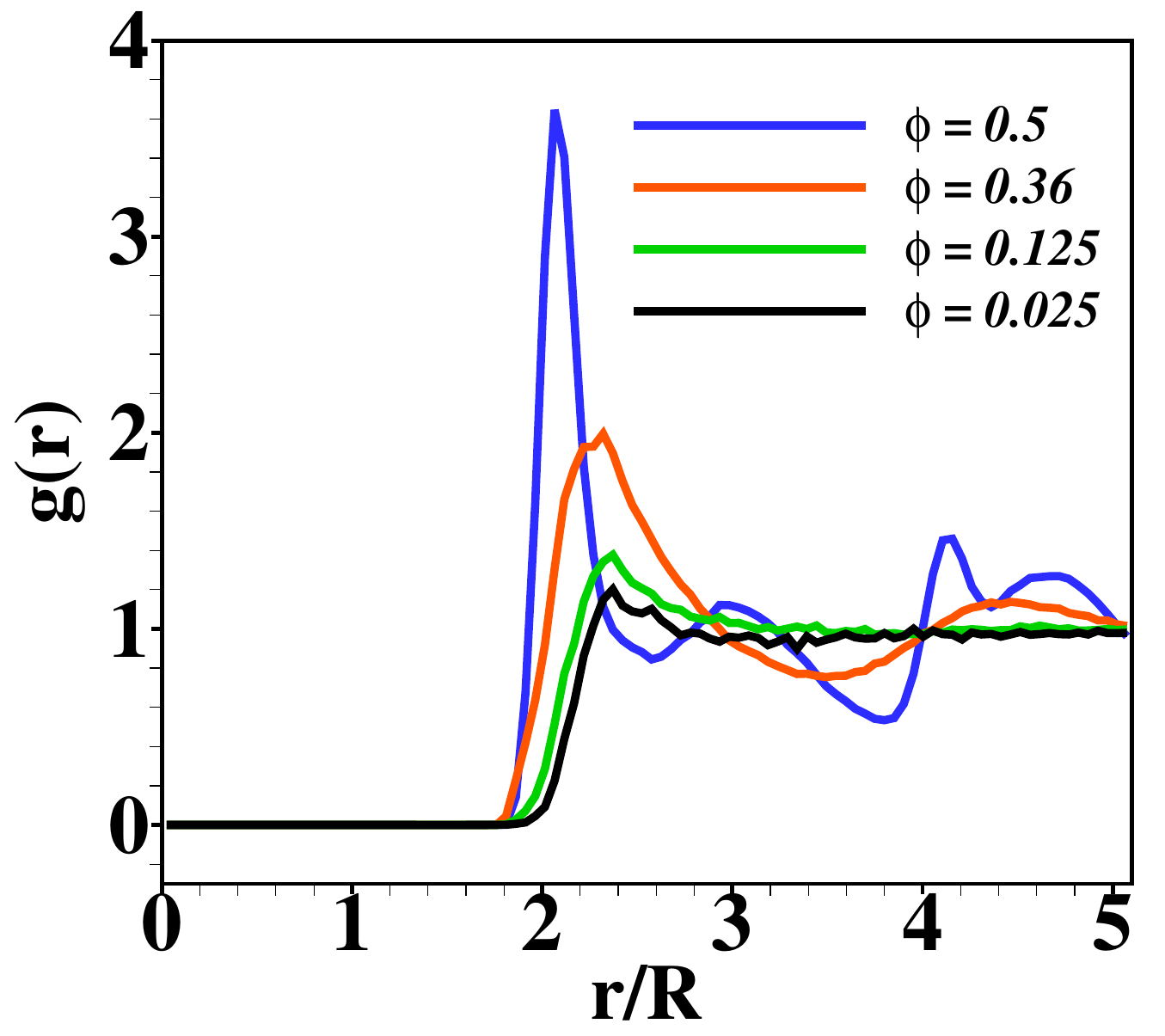}}
  \caption{The radial distribution function, $g(r)$, is obtained to characterize the microstructure of (a) spherical and (b) cubic suspensions, evaluated as a function of radial distance across various volume fractions.}
\label{RDF}
\end{figure}
The bulk volume fraction is a key factor in determining the spatial distribution and orientation of settling particles. In this section, we analyze the microstructure formation of these particles after reaching steady state. The microstructure of a particle suspension reflects the local arrangement of particles relative to each other, which is governed primarily by inter-particle hydrodynamic interactions. To characterize the overall structures within the suspension, we use the radial distribution function (RDF) \cite{sundaram1997collision,wang2000statistical}, calculated as an ensemble average over multiple independent configurations.
The RDF represents the probability of finding a particle at a distance $r$ from a reference particle, often expressed as the ratio of number density within a shell at that distance to the bulk number density, 
\begin{equation}
g(r)=\frac{n(r)}{4\pi \rho r^2 \delta r}
\label{eqRDF}
\end{equation}
Where $n(r)$ represents the number of particles within a shell of radius $r$, $\rho$ is the average particle density in the computational domain, and $\delta r$ is the shell thickness. Fig.\ \ref{RDF} displays the time-averaged RDF as a function of solid volume fraction for both suspension types. In the RDF expression given in Eq.\ \ref{eqRDF}, $g(r)$ becomes zero for distances smaller than the diameter of a particle due to steric interactions that prevent particles from overlapping, and it approaches one at larger distances as the shell density aligns with the bulk density. Fig.\ \ref{RDF} also shows that local particle ordering diminishes at shorter distances for lower volume fractions, while increasing volume fractions enhance this ordering. Peaks in the RDF indicate a higher likelihood of finding particles at specific distances, indicating the gradual formation of a particle 'cage' around each reference particle.

For dilute suspensions ($\phi \simeq 2.5\%$), the radial distribution function (RDF) exhibits a subtle peak for both spherical and cubic particles, though it is slightly lower for cubic suspensions (Fig.\ \ref{RDF}). In this regime, weak hydrodynamic interactions allow particles to form transient pairs, maintaining contact for extended periods. Consequently, a slightly higher local density of spherical particles is observed around the reference particle compared to cubic particles in this regime. As $\phi$ increases, hydrodynamic interactions among multiple particles become more pronounced, leading to the formation of aggregates or clusters in close contact, as reflected in the sharper RDF peaks. Unlike spherical particles, which maintain a fixed contact point at \(r/R = 2\) (Fig.\ \ref{RDF}a), cubic particles interact over a broader range of center-to-center distances \cite{seyed2019dynamics}. As a result, the RDF profile for cubic particles appears to be more oblique (Fig.\ \ref{RDF}b). At a moderate solid volume fraction ($\phi \simeq 36\%$), the particle-pair distribution undergoes a gradual transition. Initially, particle pairs accumulate around $r/R = 2.2$ for spheres and $r/R = 2.33$ for cubes. This enrichment later shifts to a depletion at greater distances, approximately $r/R \simeq 3.4$ for spheres and $r/R \simeq 3.5$ for cubes. The reduced inter-particle spacing compared to dilute conditions increases the likelihood of cluster formation at short distances. This effect is more pronounced for spherical particles, as evidenced by their higher RDF peaks. In contrast, cubic particles exhibit a greater tendency to escape clustering due to their higher rotational velocities, a behavior also reported by Seyed-Ahmadi and Wachs \cite{seyed2019dynamics}. With further increment in $\phi$, intensified hindered settling results from reduced inter-particle distances, leading to a more isotropic particle structure. This is reflected in the persistent oscillations beyond the first RDF peak for both suspensions at $\phi \simeq 0.5$. At larger distances from the reference particle ($r/R \geq 5.1$ for both cases), the particle distribution becomes random, with $g(r) \simeq 1$.

\begin{figure}
  \centering



  \subfigure[]{\includegraphics[width=1\textwidth]{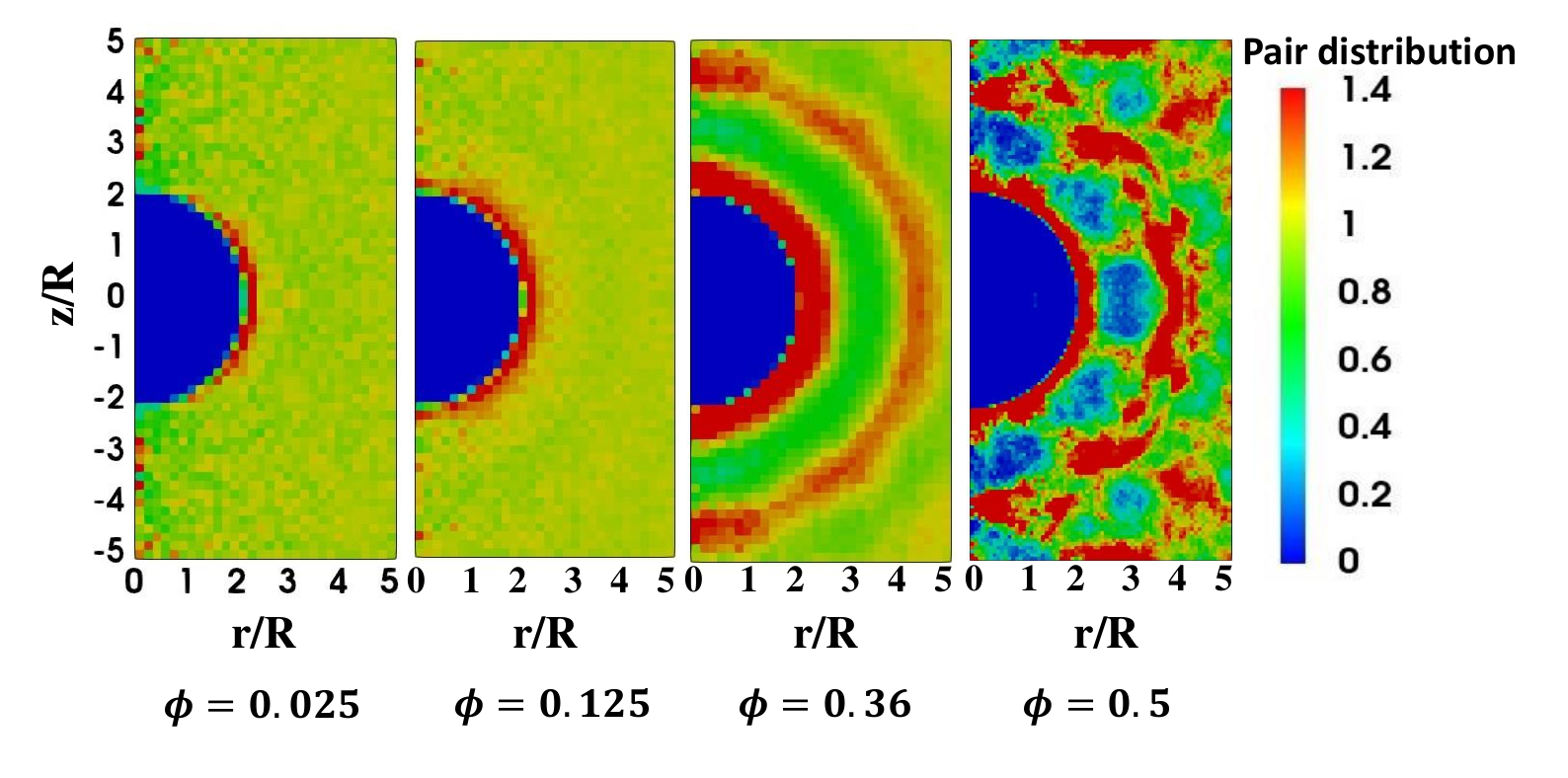}}
  \subfigure[]{\includegraphics[width=1\textwidth]{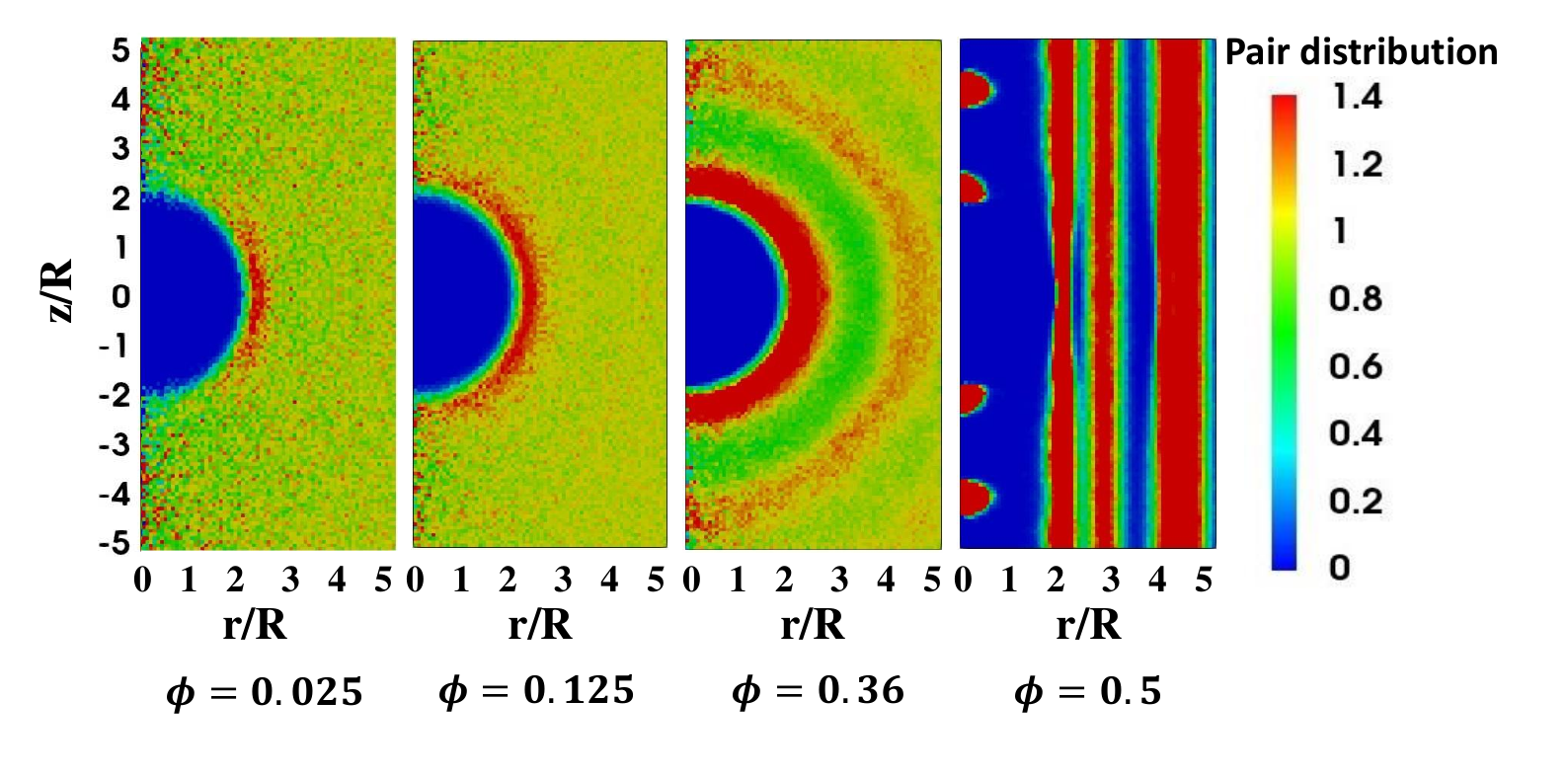}}
  \caption{The pair distribution function in cylindrical coordinates is presented for suspensions of spheres (top row) and cubes (bottom row) across varying solid particle volume fractions.}
\label{pair}
\end{figure}
Although the RDF describes the probability of finding particles within a shell regardless of orientation, it is crucial to assess whether particle pairs show directional preferences in anisotropic suspensions, such as those settling under gravity. To analyze orientation preferences, we compute the pair distribution function, $g(z,r)$, in cylindrical coordinates for both spherical and cubic suspensions. $g(z,r)$ represents the average distribution of particles located vertically at height $z$ and horizontally at distance $r$ from a reference particle. Fig.\ \ref{pair} shows the pair distribution function for various solid volume fractions.
At low volume fractions  pair-wise interactions dominate, while many particle interactions become important at higher concentrations.
As a result of inter-particle hydrodynamic interactions, settling particles naturally rearrange themselves away from their initial random configuration, leading to either clustering or dispersion.
At low volume fractions ($\phi \simeq 0.025$), there is a notable depletion of particles in the vertical direction near the reference particle for both spherical and cubic suspensions
(Fig.\ \ref{pair}). Additionally, particle pairs are enriched in the direction perpendicular to gravity, with peaks at $r/R=2.3$ for spheres and $r/R=1.5$ for cubes.
However, cubes shows a weaker concentration peak in the horizontal direction (Fig.\ \ref{RDF}).  
A similar situation is observed at $\phi=0.125$ but the anisotropy is weaker.

As the volume fraction increases, short-range hydrodynamic interactions between nearby particles become more pronounced, causing more particle pairs to cluster around the reference particle and enhancing structural ordering. We observe a transition from a preferential horizontal alignment at lower bulk concentrations ($ 0.025 \leq \phi \leq 0.125 $) to a more concentric, ring-like distribution at moderate concentrations ($ 0.15 \leq \phi \leq 0.4 $). A spherical contour of particle pairs forms at distances of $ r/R = 2.05 $ to $ 2.65 $ for spheres and $ r/R = 1.15 $ to $ 1.6 $ for cubes at $ \phi = 0.36 $ (Fig.~\ref{pair}) reveling an isotropic ordering. At this volume fraction a second ring due to second neighbors becomes visible.

The most intriguing case occurs at the highest volume fraction studied, $\phi=0.5$, where the pair distribution functions of cubes and spheres exhibit striking differences.
For spheres, the pair distribution function shows a high concentration of particles at $r/R \approx 2$ in all directions.
Beyond this region, the pair distribution function shows strong fluctuations.
As the system is very crowded spheres take longer to explore the configurational space and it is difficult to reduce the statistical errors.
In contrast, the pair distribution function for cubes shows a stronger ordering in the horizontal direction at high density.
Distinct vertical lines appear at distances around $r/R \approx 2$, followed by separations at $r/R \approx 2.7$ and $r/R \approx 4.5$,
indicating a preferred arrangement of cubes in the $xy$-plane in a square lattice with a lattice spacing $d\approx 2 R \approx 1.25 h$,
where $h=\pare{4\pi/3}^{1/3}R$ is the cube side length.
The continuous lines along the $z$-axis indicate that cubic particles lack ordering along the gravity direction.
This suggests that cubes form columnar structures capable of sliding past one another within the densely packed suspension, see also Supplementary Movie 1.

\subsection{Segregation of spherical/cubic particles in a bidisperse suspension}

Gravitational sedimentation is a fundamental technique for solid–liquid separation, widely applied in polydisperse systems where size segregation and particle sorting are essential. Although monodisperse systems, comprised of particles of uniform size, have been extensively explored \cite{brady1988stokesian,phillips1988hydrodynamic,padding2008interplay,ladd1993dynamical,guazzelli2011fluctuations,hamid2013sedimentation}, the dynamics of polydisperse suspensions (systems with particles of varying sizes, densities, or shapes) remain a topic of significant interest. Polydisperse systems present complex behavior, as the distribution of particle sizes and densities critically influences the overall flow response. Predicting the settling velocity of each particle class in a polydisperse suspension is a primary challenge; understanding these velocities is essential, as they shape the concentration profile and enable effective particle separation by size. Here, we focus on bidisperse sedimentation, which serves as a simplified yet valuable model for examining inter-particle interactions within polydisperse systems. In bidisperse suspensions with low total volume fractions, the hindrance effect becomes more complex than in monodisperse suspensions due to differential impacts of fluid displacement on the settling velocities of distinct particle classes. Although previous studies have measured settling velocities in bidisperse suspensions, these primarily focus on the settling velocities of interfaces or isolated particles within distinct zones of the suspension. When translational symmetry in the vertical direction is disrupted, for example, by the presence of horizontal boundaries, distinct concentration regions can form, leading to varied settling dynamics. In such cases, measurements are typically concentrated on the settling velocities at the interfaces between these zones \cite{lockett1979sedimentation,davies1968experimental,al1992sedimentation,chen2023characterising}. In bidisperse suspensions, where particles differ in radii, two primary sedimentation fronts generally propagate from top to bottom: one at the boundary between clear fluid and small particles, and another between small particles and a homogeneous suspension containing both particle sizes. The ratio of particle densities and sizes significantly affects size-based segregation in these systems. Numerical simulations using Stokesian dynamics have been conducted on bidisperse suspensions, covering size ratios up to 4 \cite{cunha2002modeling,wang2015short}. Recently, Li and Botto \cite{li2024hindered} examined the settling velocity statistics for dilute, non-Brownian, polydisperse sphere suspensions using Stokesian Dynamics, analyzing how total volume fraction and particle size distribution width influence sedimentation behavior. They also reported that in a polydisperse mixtures of small and large spheres, the normalized average settling velocity decreases more rapidly with increasing $\phi$ for smaller particles.

\begin{figure}[h]
  \centering
  \subfigure[]{\includegraphics[width=0.23\textwidth]{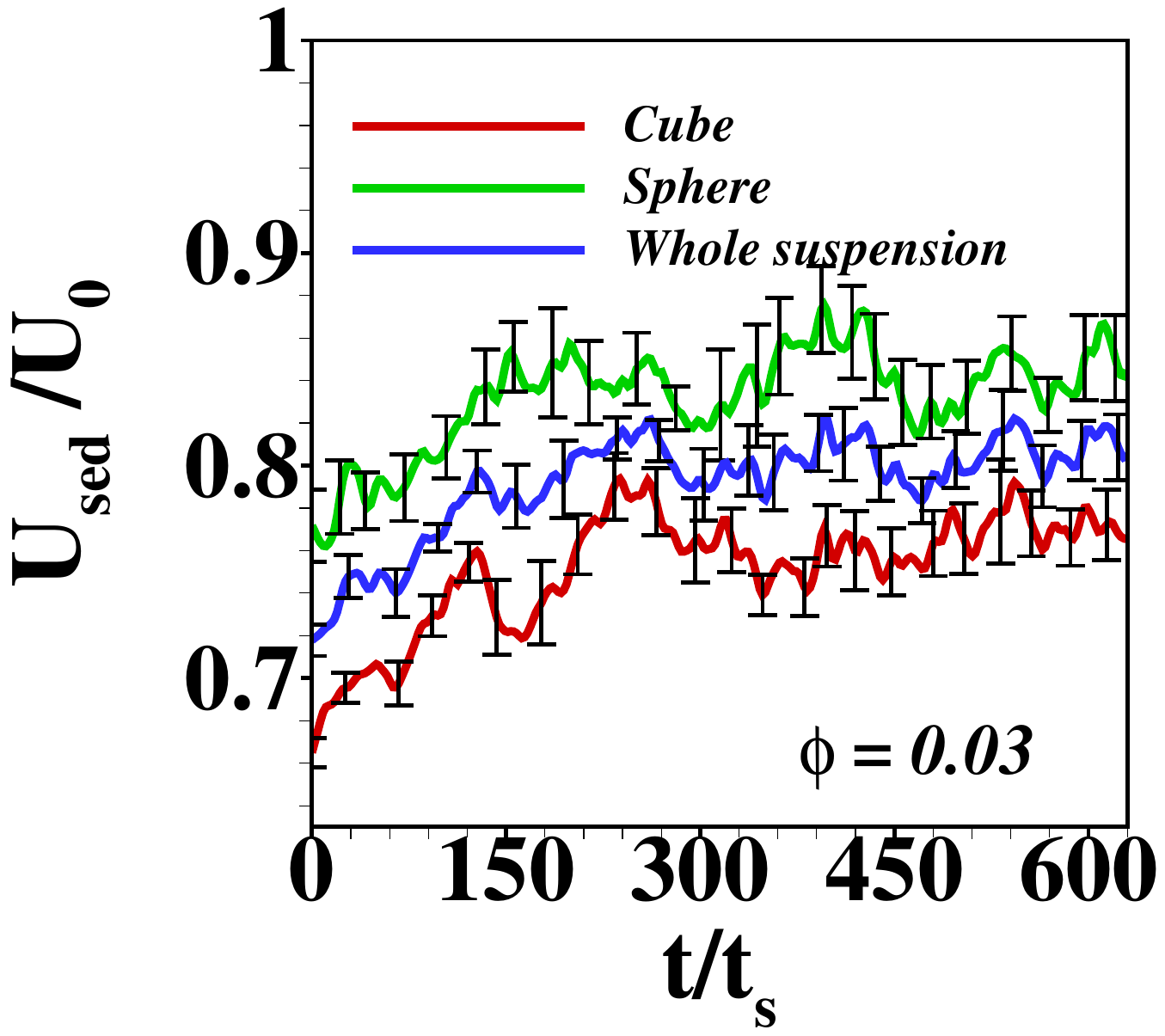}}
  \subfigure[]{\includegraphics[width=0.23\textwidth]{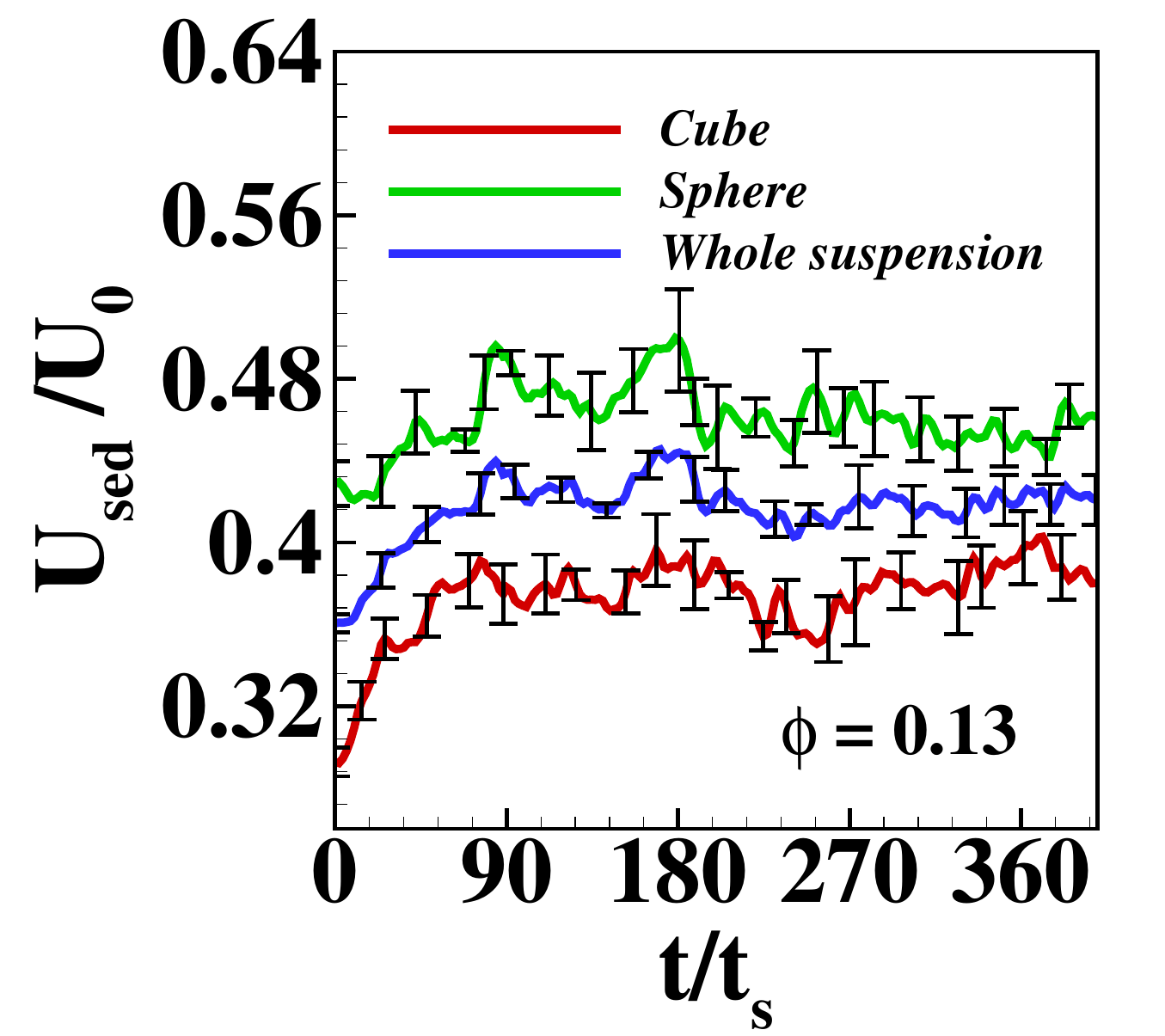}}
  \subfigure[]{\includegraphics[width=0.23\textwidth]{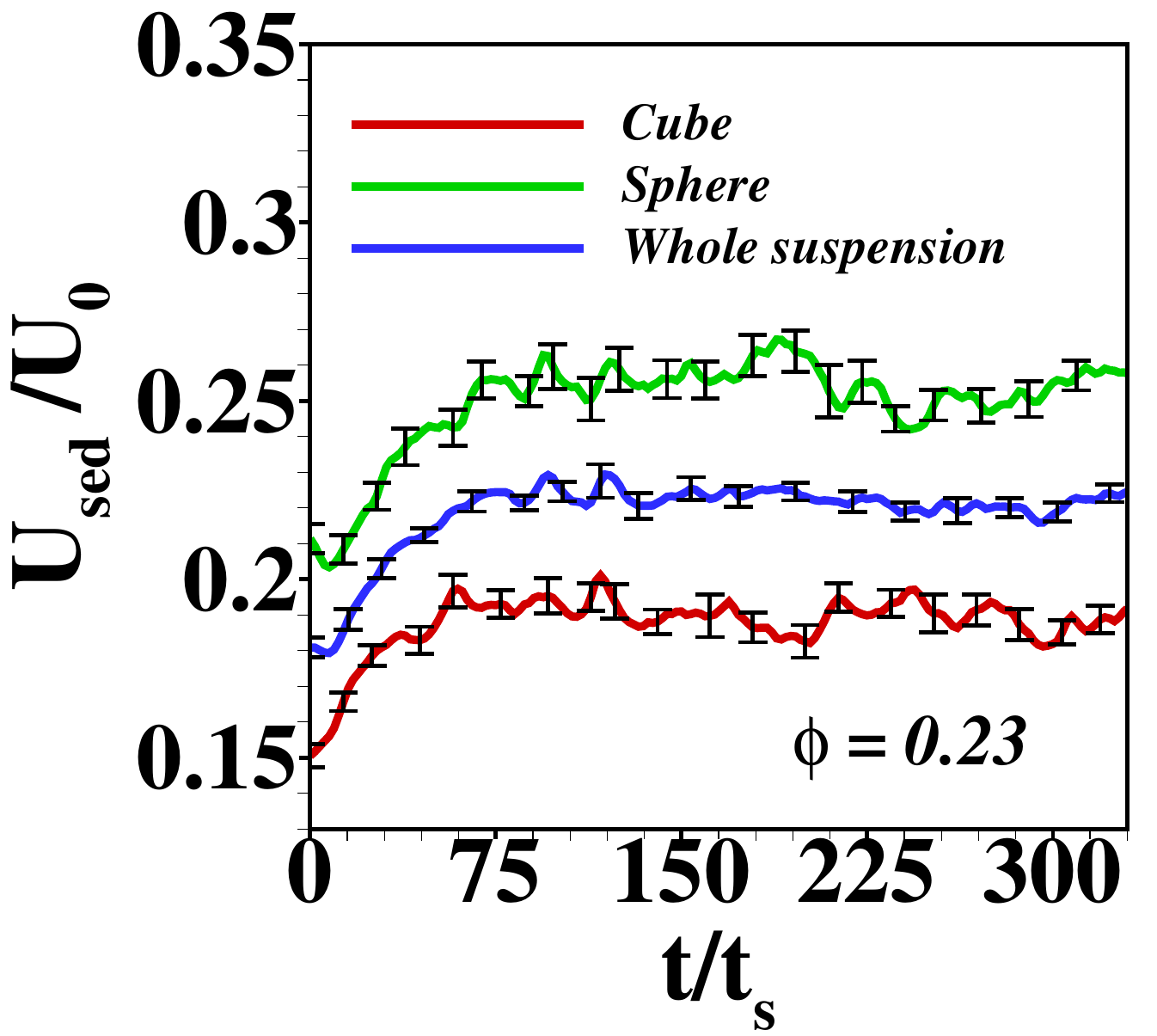}}
  \subfigure[]{\includegraphics[width=0.23\textwidth]{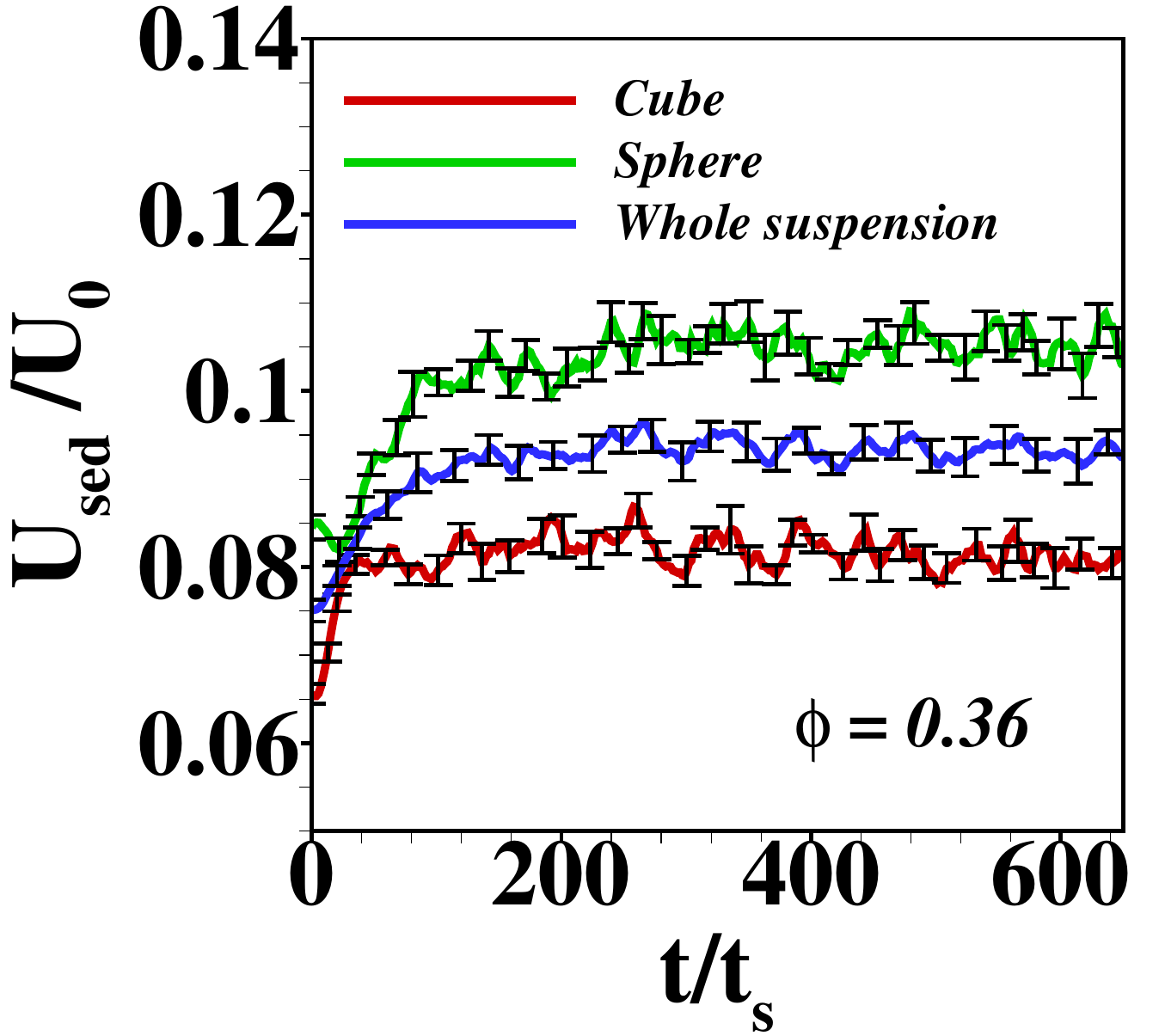}}
  \caption{Settling velocity of different species (mixture of spheres and cubes) in a bidisperse suspension is shown for different volume fraction ($\phi$) as a function of time. Velocity of each species is normalized by the Stokes settling velocity of an isolated sphere ($U_0$) in a triply periodic box. We usually employ Stokes time ($t_s=R/U_0$) as a time scale, where $R$ is taken as the radius of sphere. Average sedimentation velocity of the spheres, cubes and the whole suspension are indicated by green, red and blue solid lines, respectively.
  }
\label{mixture}  
\end{figure} 
In this section, we examine the settling velocity of non-Brownian bidisperse suspensions composed of a mixture of cubes and spheres, spanning a range of particle volume fractions from low to high. We assume all particles have a uniform mass density and identical volumes. To achieve this, we calculate an equivalent side length for the cubes, ensuring that each cube has the same volume as a sphere with a unit radius. In the bidisperse case, we consider a total of $150$ particles, evenly divided between the two particle shapes. We define the total solid volume fraction ($\phi$) as the sum of the volume fractions of each particle type: $\phi = \phi_1 + \phi_2$, where $\phi_1$ and $\phi_2$ represent the volume fractions of spheres and cubes, respectively. We examine the case of equal volume fractions for both particle types, i.e., $\phi_1 = \phi_2$,
and apply equal downward force ($f=1$) on all the particles.
As illustrated in Fig.\ \ref{mixture}, the sedimentation velocity of the bidisperse suspension is strongly influenced by the volume fraction of each species. To reduce statistical error, the velocity distributions are averaged across several random initial configurations of particles within a triply periodic simulation box. In the most dilute case ($\phi = 0.03$), the suspension requires a longer time to reach a steady state. The average sedimentation velocity of each species plays a key role in the overall settling dynamics of the suspension. As shown in Fig.\ \ref{mixture}a, spheres settle faster than cubes. This disparity is due in part to the higher angular rotation induced by the sharp edges of the cubes \cite{seyed2019dynamics,seyed2021sedimentation}, which enhances the hindrance effect relative to that experienced by spheres. As the volume fraction increases, interparticle hydrodynamic interactions become more pronounced, causing a moderate reduction in the settling velocity of each species. At higher volume fractions ($\phi = 0.36$), the suspension quickly reaches a steady state, with the reduced settling velocity attributed to increased hindrance effects and diminished velocity fluctuations in both perpendicular and parallel directions. Our study shows that spheres settle faster than cubes at all volume fractions, with segregation efficiency improving at lower $\phi$.

\begin{figure}[h]
  \centering
  \includegraphics[width=0.5\textwidth]{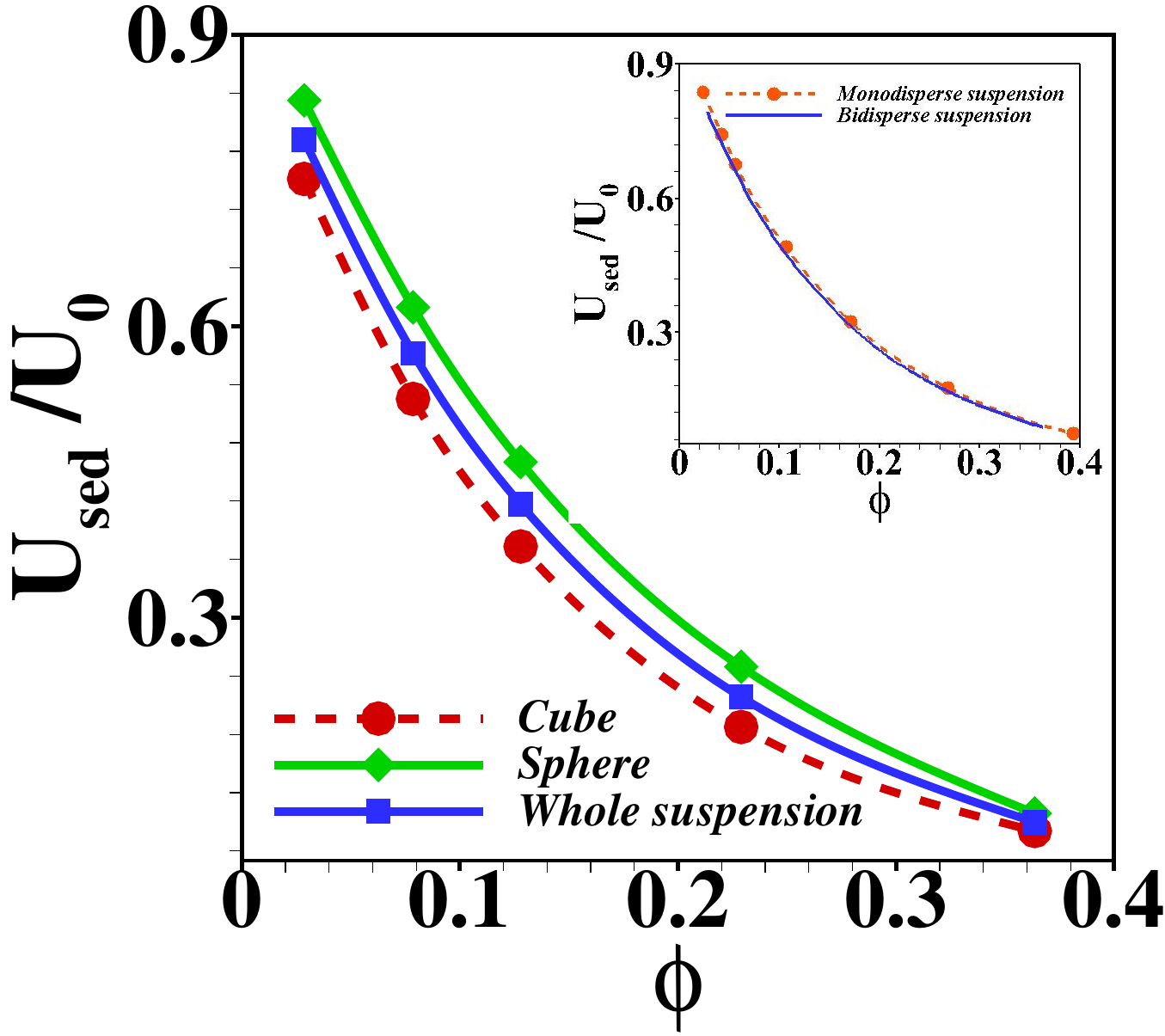}
  \caption{ The average settling velocity of the particles, $U_\text{sed}$, normalized by the Stokes velocity, $U_0$, is shown as a function of volume fraction ($\phi$). In this context, $U_0$ represents the settling velocity of an isolated sphere in the same periodic domain. Here, the error bars are smaller than the size of the symbols. Inset shows the difference of mean sedimentation velocity of the bisdisperse suspension with that of a monodisperse suspension of spheres. }
\label{mixture_sed}  
\end{figure}
Fig.\ \ref{mixture_sed} presents the normalized average settling velocity for different particle types in a bidisperse mixture across various bulk concentrations. The hindered settling behavior follows a trend similar to that observed in monodisperse suspensions (Fig.\ \ref{fig1}), but the mean settling velocity of individual species differs significantly. While the overall velocity of the mixture is slightly lower than that of a monodisperse suspension of spheres in the dilute limit, the variation in settling velocities between species is notable.
Fig.\ \ref{mixture_sed} shows that spheres settle at a rate of $11\%$ faster than cubes in this range (i.e.\ $U^{\text{sphere}}_{\text{sed}} / U^{\text{cube}}_{\text{sed}}=1.11$).
In a mixture of regular and irregular particles, spheres exhibit the most hydrodynamic shape, minimizing drag and enabling faster settling.
In contrast, irregular particles like cubes, with sharp edges and flat faces, experience higher drag, leading to slower descent.
As the volume fraction ($\phi$) increases the generated upward flux of the surrounding fluid increases which slows the settling rate.
In this regime the relative velocity difference between spheres and cubes increases to $34 \%$ at $\phi \approx 0.25$.
With further increases in $\phi$, hindered settling becomes more pronounced, reducing somewhat the segregation efficiency.
At higher concentrations ($\phi \approx 0.36$), the relative velocity difference narrows to $24 \%$,
and the overall mixture velocity approaches to a comparable magnitude that of a monodisperse suspension of spheres.

\begin{figure}
  \centering
  \subfigure[]{\includegraphics[width=0.31\textwidth]{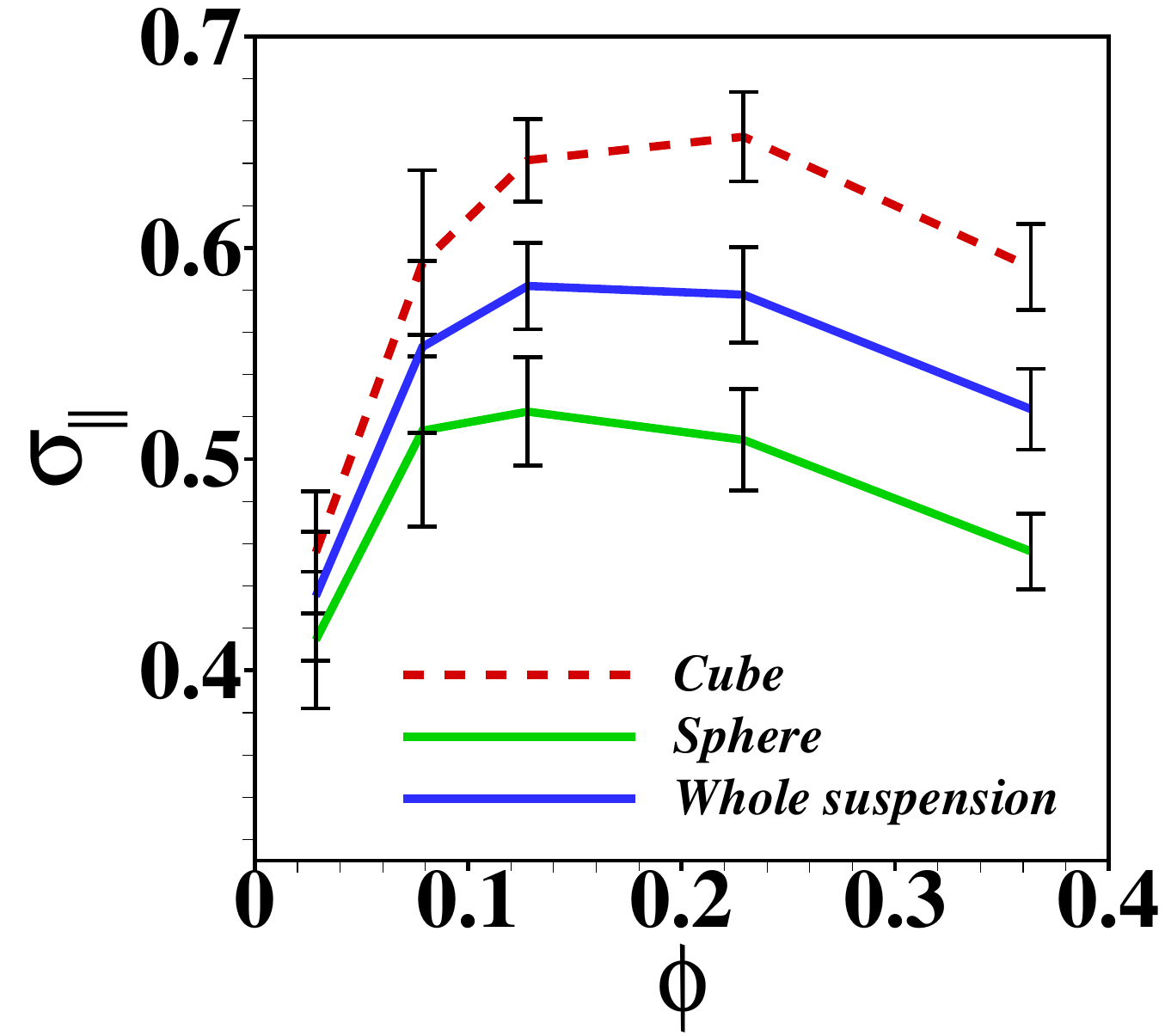}}
  \subfigure[]{\includegraphics[width=0.31\textwidth]{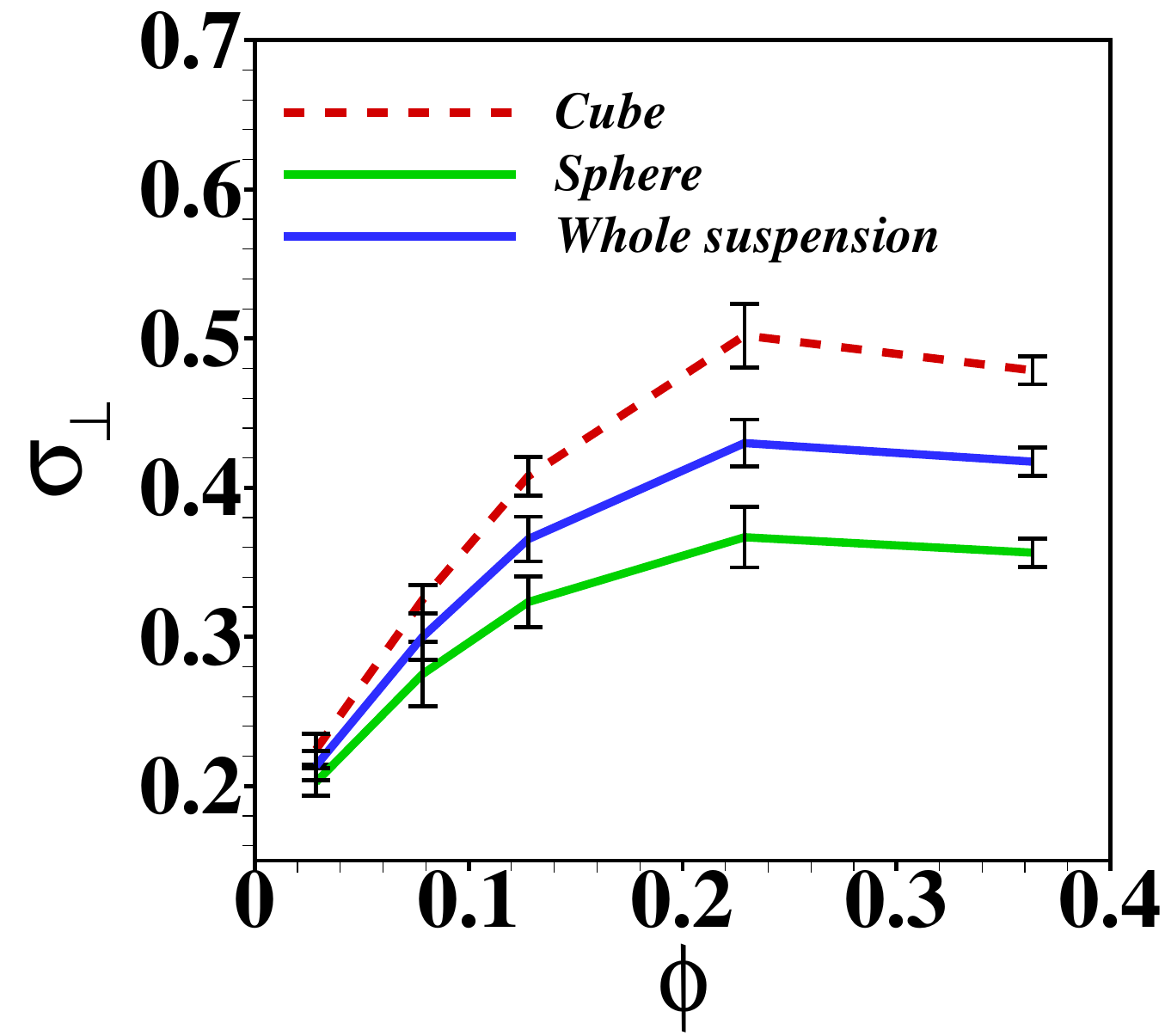}}
  \subfigure[]{\includegraphics[width=0.31\textwidth]{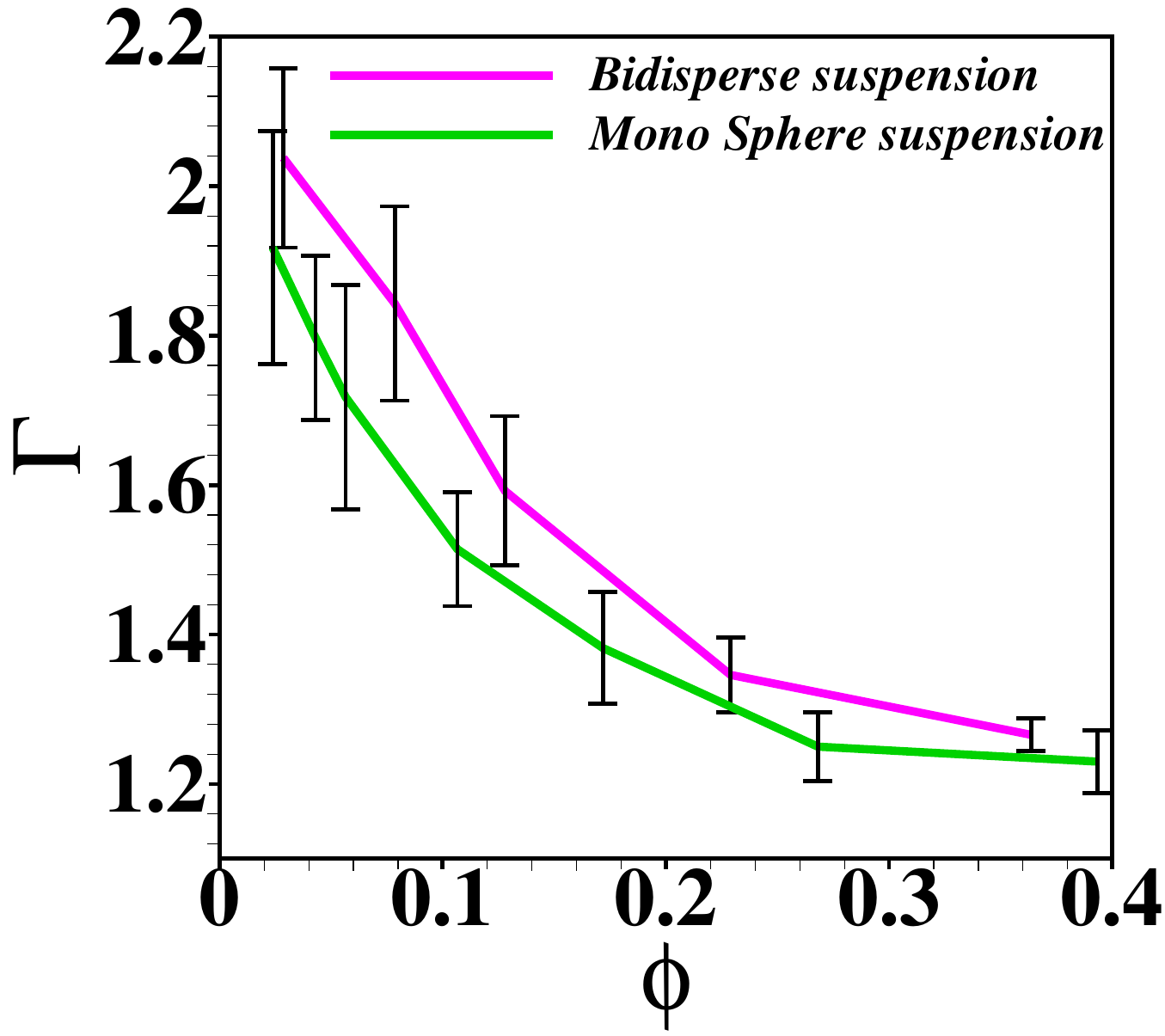}}
  \caption{Dimensionless hydrodynamic velocity fluctuations of particles as a function of volume fraction ($\phi$) are shown for (a) the settling direction ($\sigma_{\parallel}$) and (b) the direction perpendicular ($\sigma_{\perp}$) to it. (c) Comparison of anisotropy in velocity fluctuations of bidisperse mixture with the monodisperse sphere suspension. }
\label{mixture_velfluc}  
\end{figure}

In a polydisperse suspension the particles shape also affects the velocity fluctuations.
Fig.\ \ref{mixture_velfluc}a,b presents the velocity fluctuations of individual species and the entire mixture in directions parallel and perpendicular to gravity.
For dilute bidisperse suspensions ($\phi \approx 0.1$), Fig.\ \ref{mixture_velfluc}a shows a clear difference in the velocity fluctuations
in the vertical direction between the two species.
In this regime, velocity fluctuations in both directions are more pronounced than in monodisperse suspensions.
Specifically, the fluctuations of the cube velocities in the vertical direction are a $26\%$ higher in the mixture than in the monodisperse suspension
while for spheres the relative difference it is only a $4\%$.
This results in an overall rise in vertical velocity fluctuations in the bidisperse mixture compared to monodisperse cases.

As the solid volume fraction increases, reduced interparticle distances lead to more frequent collisions and stronger hydrodynamic interactions. This effect causes a rise in vertical velocity fluctuations ($\sigma_{\parallel}$) around $\phi \approx 0.23$ for both particle types. In this range, cubic particles in a bidisperse mixture show a $40 \%$ increase
in $\sigma_{\parallel}$ compared to monodisperse cubic suspensions, while spherical particles exhibit a $27 \%$ increase.
Simultaneously, horizontal velocity fluctuations steadily increase for both species.
These trends collectively alter the degree of anisotropy in velocity fluctuations, as shown in Fig \ref{mixture_velfluc}c.
The combined contributions of individual species lead to a net rise in both $\sigma_{\parallel}$ and $\sigma_{\perp}$ for the entire mixture, directly influencing the anisotropy of velocity fluctuations.
As the volume fraction increases from low to moderate values, the anisotropy parameter, $\Gamma$ decreases (Fig.\ \ref{mixture_velfluc}c).
While $\Gamma$ follows the same qualitative trend as in Fig \ref{velfluc}c, a noticeable difference emerges between the whole mixture and monodisperse sphere suspensions.
However, the absolute difference diminishes as the solid volume fraction increases due to a steady rise in horizontal fluctuations.
At higher $\phi$, intensified hydrodynamic interactions suppress velocity fluctuations in both directions, stabilizing particle motion and leading to a gradual decay in $\sigma_{\parallel}$ and $\sigma_{\perp}$ for both species.
Consequently, the overall anisotropy of the bidisperse mixture decreases, though it remains slightly higher than in the monodisperse case.  
The pair distribution function shows a weaker ordering in the bidisperse suspension compared with the monodisperse suspensions,
see Supplementary Fig.\ 2.
The weaker ordering could explain the higher velocity fluctuations observed in bidisperse suspensions.

\begin{figure}
  \centering
  \subfigure[]{\includegraphics[width=0.31\textwidth]{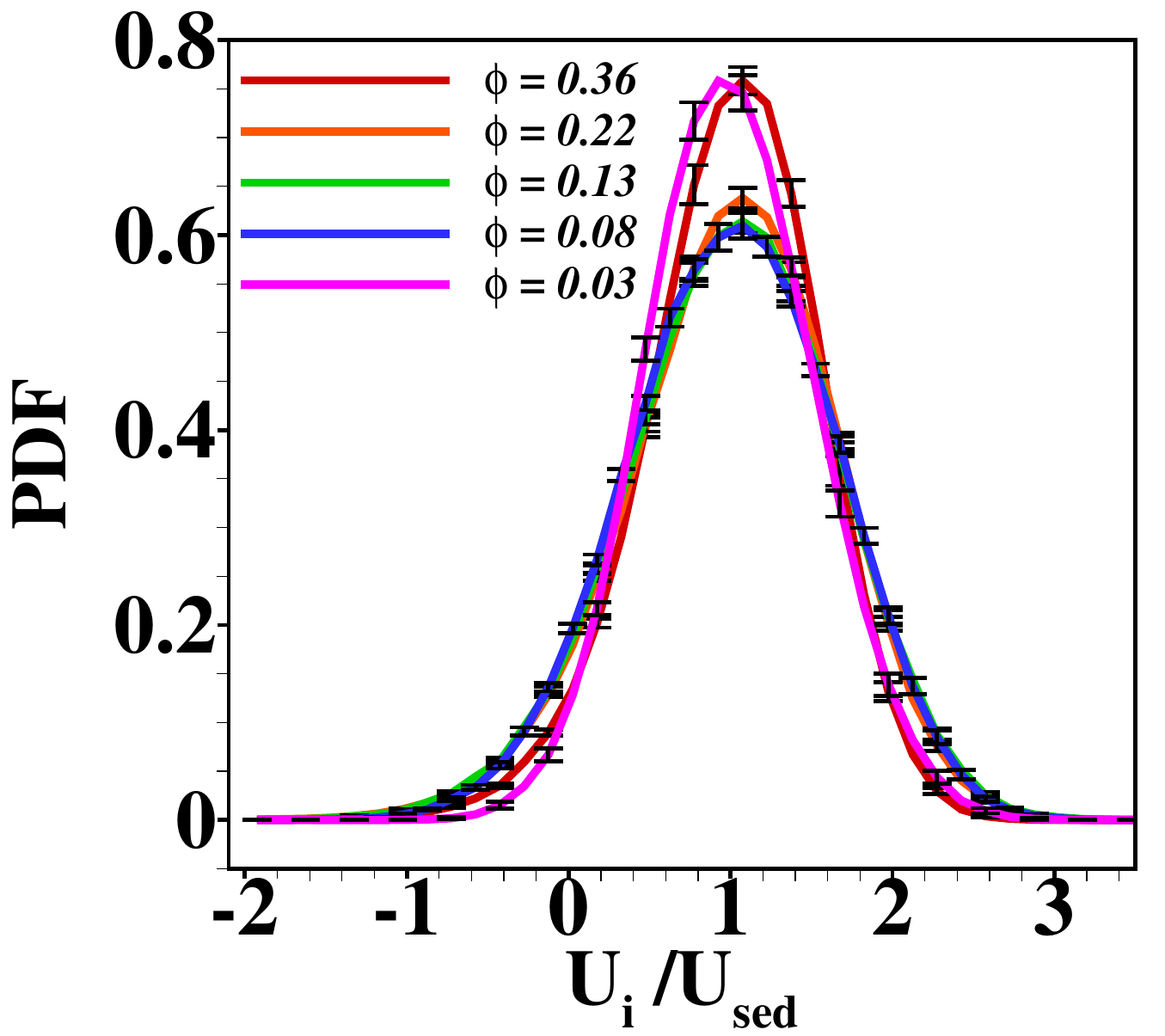}}
  \subfigure[]{\includegraphics[width=0.31\textwidth]{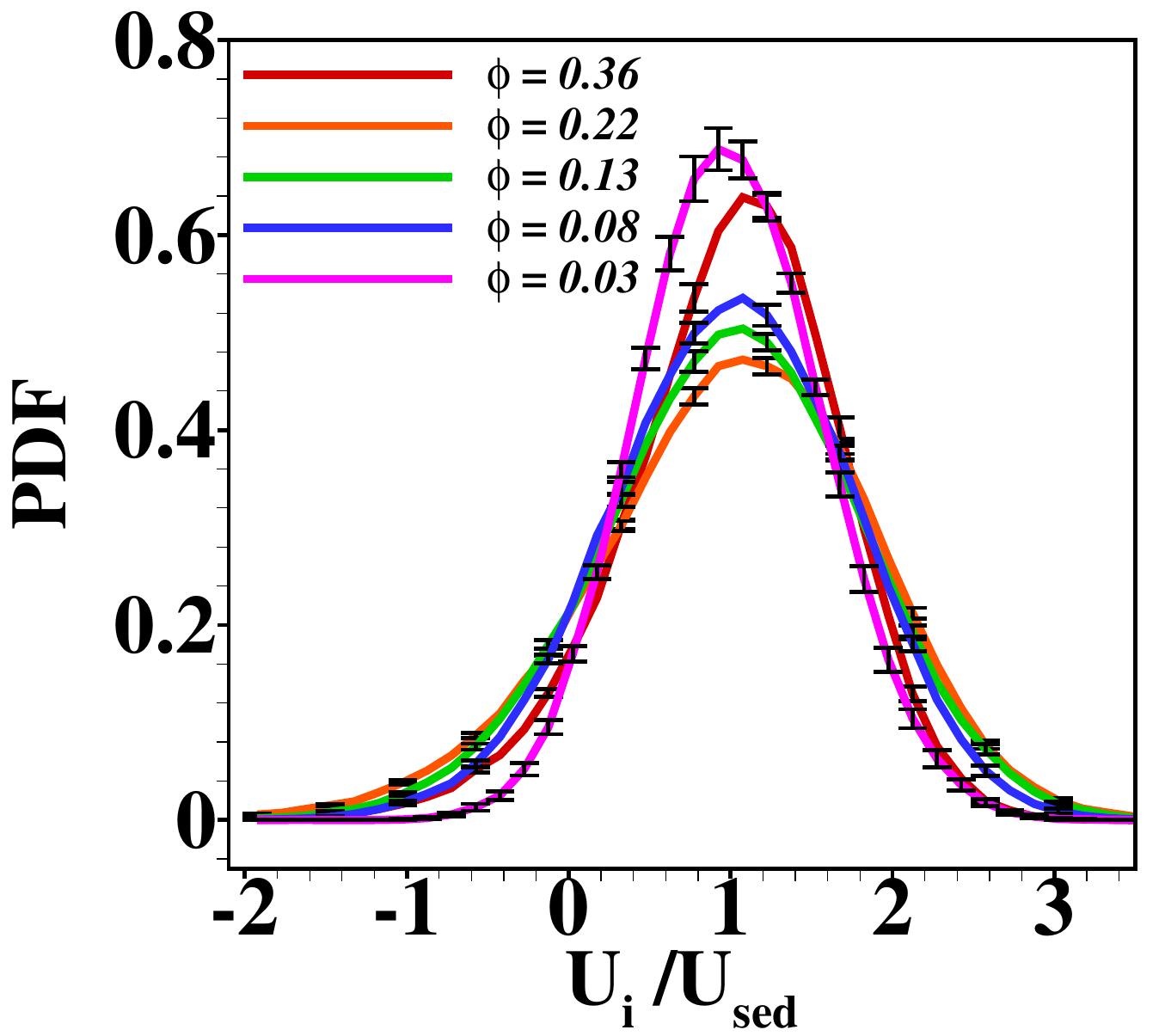}}
  \subfigure[]{\includegraphics[width=0.31\textwidth]{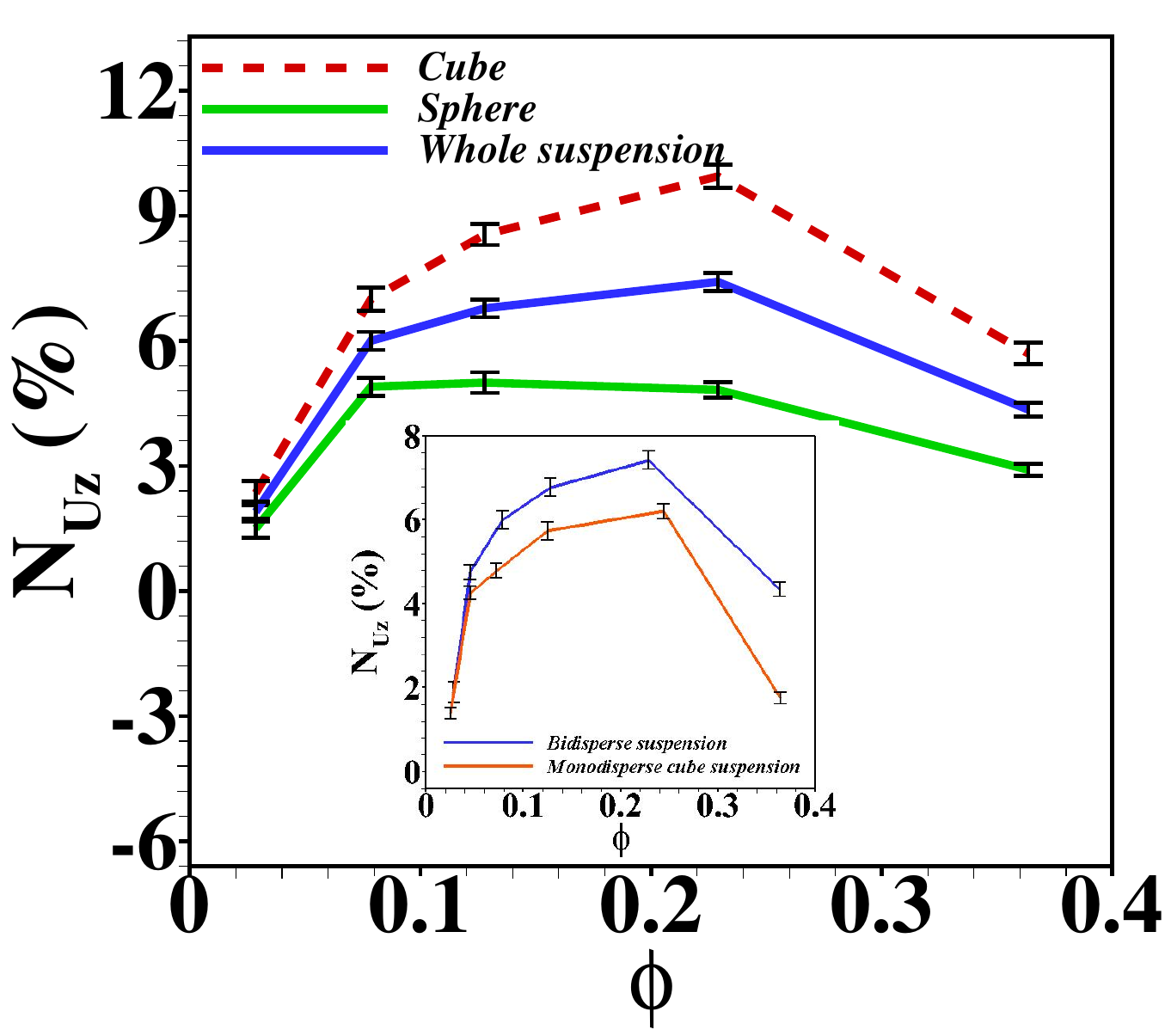}}
  \caption{ The probability density function (PDF) of the vertical velocity for (a) spherical and (b) cubic particles in a bidisperse suspension is shown across different solid volume fractions. Negative values of $ U_i / U_{\text{sed}} $ indicate particles moving upward, opposing gravity. Additionally, (c) depicts the percentage of particles ($ N_{U_z} $) moving against gravity at any given moment as a function of solid volume fraction. Inset of (c) compares $ N_{U_z} $ in a bidisperse suspension to that in a monodisperse system. }
\label{pdf_mixture}  
\end{figure}

In addition to examining the overall settling behavior and velocity fluctuations of the particle mixture, we also analyze the normalized probability density function (PDF) of vertical particle velocities (Fig.\ \ref{pdf_mixture}) for all particles within the simulation domain. While the PDF for bidisperse suspensions follows a trend similar to that of monodisperse suspensions (Fig.\ \ref{pdf}), a key distinction emerges in the distribution of vertical velocities between the two particle species in a bidisperse mixture. Across all solid volume fractions, the PDF of vertical velocity exhibits a slightly higher peak for spherical particles than for cubic particles, indicating that a larger fraction of spheres settle with velocities close to their mean sedimentation velocity. Additionally, a small fraction of particles moves instantaneously against gravity (see Supplementary Movie 2), even in the dilute regime ($\phi < 0.1$; Fig.\ \ref{pdf_mixture}a,b). The longer tail in the PDF for cubic particles suggests that cubes are more likely to exhibit upward motion than spheres, contributing to the overall velocity distribution of the mixture. As shown in Fig.\ \ref{pdf_mixture}c, $7.67 \pm 0.27 \%$ of cubic particles display upward motion, compared to $4.94 \pm 0.2 \%$ for spherical particles. This results in a total of $6.38 \pm 0.23 \%$ of particles in the bidisperse suspension moving instantaneously against gravity, higher than the $5.27 \pm 0.17 \%$ observed in monodisperse suspensions (Inset of Fig.\ \ref{pdf_mixture}c).

At moderate volume fractions, the tail of the PDF extends further into the negative velocity region due to increased velocity fluctuations and particle dispersion in the plane perpendicular to gravity. Consequently, more particles exhibit upward motion in bidisperse suspensions than in monodisperse ones. As shown in the inset of Fig.\ \ref{pdf_mixture}c, at $\phi = 0.23$, $7.39 \pm 0.2 \%$ of particles in the bidisperse suspension move upward, compared to $6.14 \pm 0.17 \%$ in the monodisperse case (see also Supplementary Movies 3 and 4). However, as $\phi$ increases further, intensified hydrodynamic interactions and inter-particle collisions suppress velocity fluctuations in both directions. As a result, at $\phi = 0.36$, only $6 \pm 0.26 \%$ of cubic particles and $3 \pm 0.14 \%$ of spherical particles exhibit upward motion, highlighting the impact of enhanced particle interactions on the settling dynamics of the mixture.

\section{Conclusion} 
\label{section 5}

This study provides a comprehensive investigation of sedimentation mechanisms in cube and sphere suspensions across a wide range of solid volume fractions, from dilute to dense regimes. Using the Rigid Multiblob (RMB) Method, we numerically predict particle dynamics within a periodic domain, uncovering key aspects of particle interactions in sedimenting systems. Our simulations align well with existing empirical correlations and experimental data for sphere suspensions when fluid inertia is neglected. Particle-particle and particle-fluid interactions play a crucial role in determining the macroscopic properties of sedimenting suspensions, with their influence strongly dependent on the global particle volume concentration. At low volume fractions, velocity fluctuations scale as $\phi^{1/2}$ due to microstructural effects but diminish at higher concentrations. Notably, horizontal velocity fluctuations are significantly higher for cubes due to stronger orientation-induced forces. Cubes also transfer momentum more effectively from gravity to the horizontal direction, as indicated by the ratio of vertical to horizontal velocity fluctuations. The anisotropy in velocity fluctuations decreases monotonically with increasing volume fraction. The strong transverse motions of cubes not only homogenize the suspension structure but also make particle momentum more isotropic compared to sphere suspensions. Additionally, in moderate to dense regimes, some particles exhibit transient upward motion against gravity. This behavior is likely driven by the formation of particle-rich and particle-poor regions, which induce local fluid convection, causing upward motion in particle-poor areas.

Microstructural analysis reveals that at low volume fractions, an anisotropic microstructure governs transport properties, with particles predominantly orienting horizontally around a test particle. This orientation effect is more pronounced for spheres than cubes, as reflected in the sharper peak of the radial distribution function (RDF) for spheres. As the volume fraction increases, many-body interactions and microstructural effects collectively control transport properties.

We also examine bidisperse suspensions composed of spherical and cubic particles, tracking the evolution of settling velocities from an initially random distribution to a steady state. In bidisperse systems, particle settling is affected by mutual interference, with settling velocities strongly dependent on solid volume fractions in the low to moderate range. Horizontal and vertical velocity fluctuations increase for each species in comparison with the monodisperse suspensions, significantly influencing suspension kinematics by enhancing anisotropy in velocity fluctuations. The presence of different particle shapes introduces inhomogeneous velocity fields due to varying drag forces, leading to differential settling speeds between species. Our results show that spheres settle faster than cubes across all volume fractions, promoting species segregation. Furthermore, increased velocity fluctuations and stronger hydrodynamic interactions in bidisperse suspensions generate locally strong return fluid fluxes, temporarily dragging more particles upward, particularly in moderate to dense regimes, highlighting the intensity of particle interactions in these conditions.

\section*{Acknowledgments}
This work has been partially funded by the Basque Government  through the projects Elkartek SosIAMet KK-2022/00110 and the BERC 2022– 2025 program. Financial support by the Spanish State Research Agency through BCAM Severo Ochoa Excellence Accreditation CEX2021-001142-S/ MICIN/AEI/10.13039/501100011033 and the Project No. PID2020-117080RB-C55 “Microscopic foundations of soft matter experiments: computational nano-hydrodynamics (Compu-Nano-Hydro)”  is also acknowledged.



\bibliographystyle{unsrt}
\bibliography{bibfile}    

\end{document}